\documentclass[prd,preprint,showpacs,preprintnumbers,amsmath,amssymb,eqsecnum,nofootinbib,groupedaddress,superscriptaddress,pdflatex]{revtex4-2}

\usepackage{amsfonts,mathrsfs}
\usepackage{bm}
\usepackage{graphicx,color}
\usepackage{hyperref}
\usepackage{ulem}
\newcommand{\void}[1]{}{}
\begin{document}
\preprint{RUP-25-3}
\title{Singularity resolution and regular black hole formation in gravitational collapse in asymptotically safe gravity}

\author{Tomohiro Harada}
\email{harada@rikkyo.ac.jp}
\affiliation{Department of Physics, Rikkyo University, Toshima, Tokyo 171-8501, Japan}
\author{Chiang-Mei Chen}
\email{cmchen@phy.ncu.edu.tw}
\affiliation{Department of Physics, National Central University, Zhongli, Taoyuan 320317, Taiwan}
\affiliation{Center for High Energy and High Field Physics (CHiP), National Central University, Zhongli, Taoyuan 320317, Taiwan}
\author{Rituparna Mandal}
\email{drimit.ritu@gmail.com}
\affiliation{Department of Physics, National Central University, Zhongli, Taoyuan 320317, Taiwan}
\begin{abstract}
We adopt an effective action inspired by asymptotically
safe gravity, in which the effective gravitational constant is parametrized as
$G(\epsilon) = G_{N} /[1 + \tilde{\omega} (G_{N}^{2} \epsilon)^{\alpha}]$, where $G_{N}$ and $\epsilon$ denote Newton's gravitational constant and the energy density of the matter field, respectively, with two dimensionless model parameters, $\tilde{\omega}$ and $\alpha$.
Within this framework, we investigate the complete gravitational collapse of a homogeneous ball of  perfect fluid and find that singularity is completely resolved for $\alpha > 1$ but not for $1/2 \le \alpha \le 1$. The case of $0 < \alpha < 1/2$ is inconsistent with asymptotic safety.
Moreover, we note that although the singularity cannot be fully resolved for $\alpha = 1$,  it is significantly weakened by quantum gravity effects.
Furthermore, we successfully construct a static exterior metric which, together with the interior solution, describes the dynamical formation of regular black holes in an asymptotically flat spacetime.
The resulting regular black hole, obtained as the final static state, contains a de Sitter core and admits a static metric fully expressible in terms of the Lerch transcendent for general cases and in elementary functions for certain values of $\alpha$, including $\alpha = 2$. 
We also discuss the formation of gravastars and the late-time evaporation process of the regular black holes. 
\end{abstract}

\date{\today}

\maketitle

\newpage

\tableofcontents

\newpage

\section{Introduction and summary}

Although general relativity (GR) is the simplest and most successful theory of gravity at large distance scales, the presence of spacetime singularities signals the breakdown of its classical description at short distance scales. This suggests the necessity of a quantum theory of gravity. However, a fundamental challenge in constructing such a theory is that GR is well known to be perturbatively nonrenormalizable, which limits its predictive power in a quantum framework.

Recently, asymptotically safe gravity~\cite{Reuter:1996cp, Souma:1999, Percacci:2017book, Reuter:2019byg} has emerged as a promising and consistent approach to quantum gravity within the framework of quantum field theory, first proposed by Weinberg~\cite{Weinberg:1980gg}. This theory is based on the existence of a nontrivial fixed point of the dimensionless coupling constants in the ultraviolet (UV) regime, ensuring that the theory remains UV complete and predictive, following the Wilsonian approach to renormalization flow. Furthermore, nonperturbative renormalization group (RG) flow equations predict that the RG trajectories of the dimensionless gravitational constant and the cosmological constant flow towards this fixed point, rendering the Einstein-Hilbert action nonperturbatively renormalizable, unlike in traditional perturbative approaches.
In the context of asymptotic safety, the existence of a nontrivial fixed point implies that Newton’s gravitational constant vanishes at high energies, leading to a weakening of gravity at such scales. This has profound implications for various phenomena in black hole physics and the resolution of singularities.

In classical GR, singularity theorems have demonstrated that spacetime singularities are inevitable in various strong gravity scenarios, including complete gravitational collapse under physically reasonable assumptions~\cite{Penrose:1964wq, Hawking:1970zqf, Hawking:1973uf, Wald:1984rg}. 
As a crucial example, Oppenheimer and Snyder~\cite{Oppenheimer:1939ue}
have shown that the complete gravitational collapse of a homogeneous dust ball
inevitably results in black hole formation 
with singularity behind an event horizon.
Conventional physics breaks down at spacetime singularities, and in quantum gravity, singularities are expected to be either resolved or at least properly addressed. If singularities are to be resolved in quantum gravity, then the final product of gravitational collapse cannot be a singular black hole but must be something else. Black holes with horizons but without singularities are referred to as regular black holes.

The first notable examples of such objects were provided in Refs.~\cite{Bardeen:1968, Hayward:2005gi}. For an assessment of the physical properties of several interesting regular black holes, see Ref.~\cite{Maeda:2021jdc} and references therein. Inspired by asymptotically safe gravity, the resolution of spacetime singularities has been explored in Refs.~\cite{Bonanno:2016dyv, Chen:2022xjk, Chen:2023wdg, Chen:2023pcv}, where the gravitational constant in the metric is replaced by an effective one. However, this substitution is somewhat ad hoc. Gravitational collapse is studied in a similar approach in Ref.~\cite{Hassannejad:2024cbu}, where the gravitational 
constant in the exterior metric is replaced by its
effective one as a function of the radial coordinate and that in Friedmann equations for the homogeneous interior is 
replaced by a function of the time coordinate.

Regular black holes typically exhibit a de Sitter core. On the other hand, highly compact objects with a de Sitter core but without horizons can also be constructed; these are known as gravastars~\cite{Mazur:2004fk, Visser:2003ge}. While early models of gravastars involved singular hypersurfaces, these are not essential ingredients. Recently, a field-theoretical construction of gravastars has been achieved, ensuring continuity in all physical quantities~\cite{Ogawa:2023ive, Ogawa:2024joy}.

Rather than arbitrarily modifying the gravitational constant in the metric, an alternative approach is to modify the coupling constants in the effective action by introducing a scale-dependent formulation and studying the effects of quantum gravity on the solutions of this effective action. One pioneering study follows the so-called Brans-Dicke approach~\cite{Reuter:2004}.
In a different framework developed by Markov and Mukhanov~\cite{Markov:1984nw}, matter-gravity coupling is introduced as a scalar function of the fluid's energy density, $\epsilon$, within the action. This approach naturally determines the effective gravitational and cosmological constants, $G = G(\epsilon)$ and $\Lambda = \Lambda(\epsilon)$, as functions of the energy density. Bonanno {\it et al.}~\cite{Bonanno:2023rzk} applied this approach within the context of asymptotically safe gravity, specifying the precise form of $G(\epsilon)$ based on the running of the gravitational constant with a physically motivated cutoff scale. Their analysis demonstrated that this mechanism resolves the spacetime singularity in the marginally bound collapse of a homogeneous dust ball. A similar approach has also been applied to cosmology~\cite{Zholdasbek:2024pxi}.
However, no consistent formation of regular black holes with a de Sitter core has been demonstrated in the context of asymptotically safe gravity, whereas such solutions have been presented in 2D dilaton gravity~\cite{Biasi:2022ktq,Barenboim:2024dko}
and in certain gravitational theories in higher dimensions~\cite{Bueno:2024eig, Bueno:2024zsx}.

In this paper, we investigate the gravitational collapse of a uniform ball of perfect fluid within the effective action framework of Ref.~\cite{Markov:1984nw}. In the absence of the first principle for identification of the energy scale $k$ with some physical scale in the considered spacetimes,
we introduce the parametrization for the effective gravitational constant as $G(\epsilon)=G_{N}/[1+\tilde{\omega} (G_{N}^{2}\epsilon)^{\alpha}]$, where $G_{N}$ is Newton's gravitational constant and $\tilde{\omega}$ and $\alpha$ are dimensionless 
positive parameters. The choice of $\alpha = 1$ corresponds to the model used in Ref.~\cite{Bonanno:2023rzk}.
We show that only for $\alpha>1$, the effective
cosmological constant $\Lambda(\epsilon)$ really approaches a constant and dominates the bare matter field as $\epsilon\to \infty$. For $\alpha>1$, 
singularity formation is completely avoided regardless of the initial conditions, whereas for \( 0<  \alpha \leq 1 \), singularity avoidance is only partial, with singularities inevitably forming in the gravitationally unbound case. 
This result sheds new light on the phenomenology as the result indicates that singularity resolution depends strongly on 
how the cutoff scale appears in the running gravitational constant within the framework of asymptotically safe gravity.
Furthermore, by smoothly matching the collapsing interior to the static exterior, we successfully construct a spacetime that describes the formation of a regular black hole possessing a de Sitter core at the center 
for gravitationally unbound and marginally bound cases, while the uniform core remains finite-sized for a gravitationally bound case.

This paper is organized as follows. In Sec.~\ref{sec:formulation}, we present the Markov-Mukhanov formulation with the parametrized effective graviational 
constant and discuss the behavior of the effective gravitational and cosmological constants.
In Sec.~\ref{sec:resolution}, we derive the dynamical equations for the Friedmann-Lema\^itre-Robertson-Walker (FLRW) spacetime and discuss the resolution of singularities there. In Sec.~\ref{sec:exterior}, we examine the metric in the exterior region and the formation of regular black holes and gravastars. In Sec.~\ref{sec:discussion}, we explore the evaporation of regular black holes and the interpretation of the exterior metric. Section~\ref{sec:conclusion} 
gives concluding remark.
In Appendix~\ref{sec:matter_conservation}, we prove (non-)conservation of the bare matter field. 
We delegate the analysis on the FLRW dynamics for some cases to Appendix~\ref{sec:alphal1}. In Appendix~\ref{sec:junction}, we derive the junction conditions.
Throughout this paper, we use the sign convention of Wald~\cite{Wald:1984rg} and units where $c = \hbar = k_{B} = 1$. Newton’s gravitational constant or the Planck mass, $G_{N} = m_{P}^{-2}$, is retained explicitly.

\section{Effective action and gravitational constant \label{sec:formulation}}

\subsection{Markov-Mukhanov formulation of the effective action \label{sec:MM}}

The action for an isentropic perfect fluid in Einstein gravity can be written in Ref.~\cite{Hawking:1973uf} as
\begin{equation}
 S = \frac{1}{16\pi G_{N}} \int d^{4}x \sqrt{-g} [R - 16 \pi G_{N} \epsilon].
\label{eq:original_action}
\end{equation}
In this action, $\epsilon$ is a function of $n$, i.e., $\epsilon = \epsilon(n)$, where $n$ is the conserved number density. The number conservation 
$\nabla_{\mu} (n u^{\mu}) = 0$ is required with $u^{\mu}$ being 
the four-velocity of the fluid.
The function $\epsilon(n)$ specifies the equation of state (EOS) through the first law
\begin{equation}
 p = n \frac{d\epsilon}{dn} - \epsilon,
 \label{eq:1st_law_bare}
\end{equation}
which is derived by the variational principle. 
For example, the EOS $p = w \epsilon$
is realized by the choice $\epsilon = C n^{1+w}$, where $C$ is a nonzero constant. The variation of the action with respect to the metric gives the Einstein equation
\begin{equation}
 G_{\mu\nu} = 8 \pi G_{N} T_{\mu\nu},
\end{equation}
where the stress-energy tensor is derived as
\begin{equation}
 T_{\mu\nu} = (\epsilon + p) u_{\mu} u_{\nu} + p g_{\mu\nu},
 \label{eq:bare_matter_stress-energy_tensor}
\end{equation}
which satisfies the conservation law
\begin{equation}
 \nabla_{\mu} T^{\mu\nu} = 0.
 \label{eq:bare_matter_conservation}
\end{equation}

To discuss the effect of quantum gravity to the cosmological expansion in the early Universe, 
Markov and Mukhanov~\cite{Markov:1984nw} introduced an effective action 
\begin{equation}
 S = \frac{1}{16 \pi G_{N}} \int d^{4}x \sqrt{-g}[R - 2 \chi(\epsilon) \epsilon],
\label{eq:MM_action}
\end{equation}
where again $\epsilon=\epsilon(n)$ is assumed. 
We call $\epsilon$ and $p$, the latter of which is defined through Eq.~(\ref{eq:1st_law_bare}), the bare energy density and pressure, respectively.
 Although the suffix $B$ may be put to $\epsilon$ and $p$ to distinguish them from the effective quantities defined below, it is omitted here to avoid complicated notations except that we will cover it in the discussion in Appendix~\ref{sec:matter_conservation}.
The function $\chi(\epsilon)$ denotes the quantum correction.
If $\chi(\epsilon) = 8 \pi G_{N}$, then we recover the original action~(\ref{eq:original_action}). On the other hand, if we put $\chi(\epsilon) \epsilon = 8 \pi G_{N} \epsilon_{\rm eff}$, then everything goes in the same way as the original system except for $\epsilon$ and $p$
being replaced by $\epsilon_{\rm eff}$ and $p_{\rm eff}$. 
The first law reduces to
\begin{equation}
 p_{\rm eff} = n \frac{d\epsilon_{\rm eff}}{dn} -\epsilon_{\rm eff}.
\end{equation} 
This gives the relation between $p_{\rm eff}$
and $\epsilon_{\rm eff}$ parametrized by $n$, i.e., the EOS for the effective fluid. So, given a function $\epsilon(n)$, introducing $\chi(\epsilon)$ 
is nothing but modifying the EOS from the original one $p = p(\epsilon)$ to the effective one $p_{\rm eff} = p_{\rm eff}(\epsilon_{\rm eff})$ from a classical point of view.

The variation with respect to the metric yields
\begin{equation}
 G_{\mu\nu} = 8 \pi G_{N} T_{{\rm eff}\mu\nu},
 \label{eq:modified_Einstein_eq}
\end{equation} 
where the effective stress-energy tensor is given by 
\begin{eqnarray}
T_{{\rm eff}\mu\nu} =
 (\epsilon_{\rm eff} + p_{\rm eff}) u_{\mu} u_{\nu} + p_{\rm eff} g_{\mu\nu}
 \label{eq:effective_stress-energy_tensor}
\end{eqnarray}
with the effective matter quantities being
\begin{eqnarray}
 8 \pi G_{N} \epsilon_{\rm eff} &=& \chi \epsilon \label{eq:effective_energy_density}, \\ 
 8 \pi G_{N} p_{\rm eff} &=& p \frac{d(\chi \epsilon)}{d\epsilon} + \frac{d\chi}{d\epsilon} \epsilon^{2}.
 \label{eq:effective_pressure}
\end{eqnarray}
As we can see in Eq.~(\ref{eq:modified_Einstein_eq}), it is not the bare energy density and pressure but the effective ones that source the gravitational fields through the Einstein equation.
The effective gravitational constant $G(\epsilon)$ and cosmological constant $\Lambda(\epsilon)$ are defined as 
\begin{eqnarray}
8 \pi G_{N} T_{{\rm eff}\mu\nu} = 8 \pi G(\epsilon)[(\epsilon + p) u_{\mu} u_{\nu} + p g_{\mu\nu}] -\Lambda(\epsilon) g_{\mu\nu}, 
\label{eq:effective_bare_stress-energy_tensor}
\end{eqnarray}
so that we can find 
\begin{eqnarray}
8 \pi G(\epsilon) = \frac{d(\chi(\epsilon) \epsilon)}{d\epsilon}, \qquad \Lambda(\epsilon) = -\frac{d\chi(\epsilon)}{d\epsilon}\epsilon^{2} = 8 \pi G_{N} \epsilon_{\rm eff} - 8 \pi G(\epsilon) \epsilon.
\label{eq:Gepsilon_Lambdaepsilon}
\end{eqnarray}
Inversely, we can uniquely reconstruct $\epsilon_{\rm eff}$ and $p_{\rm eff}$ as 
\begin{eqnarray}
 8 \pi G_{N} \epsilon_{\rm eff} = 8 \pi \int^{\epsilon}_{0} G(s) ds, \qquad 8 \pi G_{N} p_{\rm eff} = 8 \pi G(\epsilon) (\epsilon + p) - 8 \pi G_{N} \epsilon_{\rm eff}.
\label{eq:effective_pressure_1}
\end{eqnarray}
where the integration constant is chosen so that $\epsilon_{\rm eff} = \epsilon$ is recovered for $G(\epsilon) = G_{N}$.
The invariance of the action against infinitesimally small coordinate transformation yields the matter conservation law for the effective matter field 
\begin{equation}
 \nabla_{\mu} T_{\rm eff}^{\mu\nu} = 0.
 \label{eq:effective_matter_conservation}
\end{equation}

On the other hand, under the action~(\ref{eq:MM_action}), the bare matter field 
stress-energy tensor defined by Eq.~(\ref{eq:bare_matter_stress-energy_tensor}) is not required to satisfy the conservation law 
(\ref{eq:bare_matter_conservation}) in general. 
However, surprisingly, in the FLRW spacetime, the conservation law for the bare matter field happens to be satisfied due to the high symmetry of the
spacetime. We will prove this in Appendix~\ref{sec:matter_conservation}.

\subsection{Effective gravitational constant inspired by asymptotically safe gravity \label{sec:Geff}}
In the framework of asymptotically safe gravity~\cite{Reuter:1996cp}, the central element is the gravitational effective average action (EAA), a coarse-grained functional of the metric and a momentum scale $k$, which acts as an infrared cut off. The construction of the EAA involves integrating out all quantum fluctuations with momenta $q^{2} > k^{2}$, while suppressing contributions from modes with $q^{2} < k^{2}$. The evolution of the effective average action is governed by the exact functional renormalization equation, which can be solved by truncating the infinite-dimensional action space into the Einstein-Hilbert truncation. Then, the effective gravitational constant and cosmological constant become scale dependent, denoted as $G(k)$ and $\Lambda(k)$, respectively. In the UV limit, the RG flow approaches a fixed point, where the dimensionless coupling constants attain finite limit values $g(k) := k^{2} G(k) \to g_{*}$ and $\lambda(k) := k^{-2} \Lambda(k) \to \lambda_{*}$ as $k \to \infty$.	
For a small value of the dimensionless cosmological constant, the RG equations lead to the solution for the Newton coupling, $G(k) ={G_{N}}/{(1+\omega G_{N}k^{2})}$, where $\omega$ is a certain positive value of the order of unity~\cite{Bonanno:2000ep}.

To include the above analysis in the spacetime dynamics, the cutoff scale $k$ can be regarded as a function of some physical quantities such as the spacetime curvature scales and some physical length scales. As long as the authors are aware, no first principle has been established
for the scale identification so far, although 
there are two major proposals, the ``curvature invariant'' proposal and 
the ``proper distance'' proposal, as we will discuss below.
In the FLRW spacetime, the local curvature scale is completely characterized by the energy density if we fix the EOS.
So, it is natural to assume that the cutoff scale is a function of the bare energy density $\epsilon$, i.e., $k = k(\epsilon)$, and identify the scale dependent coupling constants $G(k(\epsilon))$ and $\Lambda(k(\epsilon))$ with $G(\epsilon)$ and $\Lambda(\epsilon)$ in the Markov-Mukhanov formulation.
However, there is no prior relationship between $(G(k), \Lambda(k))$ and $(G(\epsilon), \Lambda(\epsilon))$.
In this situation, we introduce the parametrization for the effective gravitational constant by two positive dimensionless 
parameters $\tilde{\omega}$ and $\alpha$ such that 
\begin{eqnarray}
 G(\epsilon) 
= \frac{G_{N}}{1 + \tilde{\omega} (G_{N}^{2} \epsilon)^{\alpha}}.
\label{eq:G_epsilon_parameterise}
\end{eqnarray}
In other words, we infer the following {identification of the cutoff scale $k$ as
\begin{equation}
\omega G_{N} k^{2} = \tilde{\omega} (G_{N}^{2} \epsilon)^{\alpha}.
\label{eq:k_epsilon}
\end{equation}
A larger $\tilde{\omega}$ implies a lower quantum gravity scale in terms of $\epsilon$ at which quantum gravity effects begin to affect the system, while a larger $\alpha$ gives stronger quantum-gravity effects for $\epsilon \to \infty$.
The dimensionless constant $\tilde{\omega}$ is not necessarily of the order of unity.

It is not so straightforward to identify the value of the index $\alpha$.
We should note that $\alpha = 1$ is chosen for the FLRW spacetime with dust in Ref.~\cite{Bonanno:2023rzk} by an argument on the proper distance to the center of the classical Schwarzschild black hole in Ref.~\cite{Bonanno:2000ep}.~\footnote{In Ref.~\cite{Hassannejad:2025maw}, $\alpha=1/(1+w)$ is suggested for the equation of state $p=w\epsilon$ according to the same discussion as in Ref.~\cite{Bonanno:2023rzk}.}
If we use the scalar curvature polynomials such as the Kretschmann invariant ${\cal K}=R^{\mu\nu\rho\sigma} R_{\mu\nu\rho\sigma}$ for scale identification, we find $k \sim {\cal K}^{1/4} \sim (G_{N} \epsilon)^{1/2}$ in classical dynamics, so that we reach $\alpha = 1$.
On the other hand, a simplistic dimensional argument 
might imply $\alpha = 1/2$ because $\epsilon$ has mass dimension of $4$, in which the relation~(\ref{eq:k_epsilon}) need not involve $G_{N}$. 
If we consider radiation fluid in thermal equilibrium, which is regarded as $w=1/3$, its associated temperature $T$ is proportional to $\epsilon^{1/4}$,  naturally implying $\alpha=1/2$ together with the assumption $k\sim T$.
This seems to be consistent with the choice 
in Ref.~\cite{Bonanno:2016dyv}.
It is also interesting that the coarse graining 
scale $k = \sigma H$ is naturally 
introduced in the framework of stochastic inflation
in terms of the physical proper wave number $k$ with $\sigma \sim 0.1$, where $H$ is the Hubble parameter~\cite{Starobinsky:1982ee, Starobinsky:1986fx}.
If we identify this coarse graining scale with that in asymptotically safe gravity, $\alpha = 1$ will again be preferred because $k = \sigma H \sim \sigma (G_{N} \epsilon)^{1/2}$ using the Einstein equation. 
Furthermore, we may have to consider that the modified physical spacetime is governed by not the bare quantities $\epsilon$ and $p$ 
but the effective ones $\epsilon_{\rm eff}$ and $p_{\rm eff}$. 
In fact, later in this paper, we will discuss that $\alpha > 1$ is favored from a singularity resolution point of view.
Incidentally, although we cannot expect that the simple functional form given in Eq.~(\ref{eq:G_epsilon_parameterise}) 
should hold across all the energy scales $0 < \epsilon < \infty$, physical results about singularity resolution discussed in this paper will not depend on the detailed functional form but on the asymptotic behavior in the limit of $\epsilon \to \infty$.

In the framework of Markov and Mukhanov for a given energy density $\epsilon = \epsilon(n)$, if we fix the effective gravitational constant $G(\epsilon)$, we will obtain the effective cosmological constant $\Lambda(\epsilon)$ by Eq.~(\ref{eq:Gepsilon_Lambdaepsilon}).
The dimensionless cosmological constant 
\begin{equation}
\lambda(k) = \frac{\Lambda(k)}{k^{2}} = \frac{\omega G_{N} \Lambda(\epsilon)}{\tilde{\omega}
(G_{N}^{2}\epsilon)^{\alpha}}
\end{equation}
must have a finite limit for the consistency with the RG flow giving asymptotic safety.
So, this can be regarded as an additional condition for the physical effective gravitational constant $G(\epsilon)$.

\subsection{Effective gravitational and 
cosmological constants and perfect fluid}

Now that we have the functional form of $G(\epsilon)$ as in Eq.~(\ref{eq:G_epsilon_parameterise}), we can obtain $\epsilon_{\rm eff}(\epsilon)$, $p_{\rm eff}(\epsilon)$ and $\Lambda(\epsilon)$ from Eqs.~(\ref{eq:Gepsilon_Lambdaepsilon}), 
(\ref{eq:effective_pressure_1}) and the EOS $p=p(\epsilon)$.
The effective energy density 
$\epsilon_{\rm eff}(\epsilon)$ is given by 
 \begin{eqnarray}
 G_{N}^{2} \epsilon_{\rm eff} = G_{N} \int^{\epsilon}_{0} G(s) ds
= \tilde{\omega}^{-1/\alpha} I_{\alpha}(x), \label{eq:epsilon_eff}
\end{eqnarray}
where we have defined
\begin{eqnarray}
 I_{\alpha}(x) := \int^{x}_{0} \frac{d\tilde{x}}{1 + \tilde{x}^{\alpha}}, 
\label{eq:integral}
\end{eqnarray}
with $x = \tilde{\omega}^{1/\alpha} G_{N}^{2} \epsilon$.
The integral can be implemented as 
\begin{equation}
 I_{\alpha}(x)=x ~_{2}F_{1}\left(1, \frac{1}{\alpha}; 1+\frac{1}{\alpha}; -x^{\alpha}\right)
= \frac{x}{\alpha} \Phi\left( -x^{\alpha}, 1, \frac{1}{\alpha} \right); \quad 
 \Phi(z, s, \alpha) = \sum_{n=0}^{\infty} \frac{z^{n}}{(n+\alpha)^{s}},
\label{eq:Ialpha_hypergeometric}
\end{equation}
where $\Phi$ is called the Lerch transcendent.
For example, for $\alpha = 1/2$, $1$ and $2$, this can be expressed by elementary functions as follows:
\begin{eqnarray}
I_{1/2}(x) = 2 [ \sqrt{x} - \ln(\sqrt{x} + 1) ], \quad 
I_{1}(x) = \ln(1 + x), \quad 
I_{2}(x) = \arctan(x).
\label{eq:Ihalf_I1_I2}
\end{eqnarray}
For any positive value of $\alpha$, for $0 < x \ll 1$, we find 
$I_{\alpha}(x) \approx x$.
For $x \gg 1$, for $0 < \alpha < 1$, the integral can be estimated as 
$I_{\alpha}(x) \approx x^{1-\alpha}/(1-\alpha)$,
while for $\alpha > 1$, the integral converges as $x \to \infty$ to
\begin{equation}
 \lim_{x \to \infty} I_{\alpha}(x) = \int_{0}^{\infty} \frac{1}{1+\tilde{x}^{\alpha}} d\tilde{x} =: {\mathscr C}_{\alpha}
\end{equation}
with $0<{\mathscr C}_{\alpha}<\infty$.

We can derive the asymptotic behaviors of $\epsilon_{\rm eff}(\epsilon)$ and $\Lambda(\epsilon)$. For $\alpha>1$, we find as $\epsilon\to \infty$,
\begin{eqnarray}
 \epsilon_{\rm eff} &\to&  \epsilon_{{\rm eff}{*}} := G_{N}^{-2} \tilde{\omega}^{-1/\alpha}  {\mathscr C}_{\alpha}, 
\label{eq:eeff0}\\
 \Lambda &\to& \Lambda_{{*}} := 8 \pi G_{N} \epsilon_{{\rm eff}{*}}, \label{eq:Lambda_eff_a>1}
 \end{eqnarray}
 so both $\epsilon_{\rm eff}$ and $\Lambda$ approach finite positive limits.
 For $\alpha=1$, we find 
\begin{eqnarray}
\epsilon_{\rm eff}&=& G_{N}^{-2}\tilde{\omega}^{-1}
\ln\left(1+\tilde{\omega}G_{N}^{2}\epsilon\right), \\
 \Lambda &=& 8 \pi G_{N}^{-1} \tilde{\omega}^{-1} \left[
\ln\left( 1 + \tilde{\omega}G_{N}^{2} \epsilon \right) -\frac{\tilde{\omega} G_{N}^{2} \epsilon}{1 + \tilde{\omega}G_{N}^{2} \epsilon} \right],
\end{eqnarray}
so both show logarithmic divergence as $\epsilon\to \infty$.
For $0<\alpha<1$, we have as $\epsilon\to \infty$,
\begin{eqnarray}
\epsilon_{\rm eff} &\approx&
\frac{1}{1 - \alpha}G_{N}^{-2} \tilde{\omega}^{-1}(G_{N}^{2} \epsilon)^{1-\alpha}, \\
 \Lambda &\approx& 8 \pi G_{N}^{-1} \frac{\alpha}{1 -\alpha} \tilde{\omega}^{-1} (G_{N}^{2} \epsilon)^{1-\alpha}, 
\end{eqnarray}
so both blow up to infinity as $\epsilon\to \infty$.
It is easy to show the nondimensional cosmological constant $\lambda(k)\to 0$ as $\epsilon\to \infty$ for $\alpha\ge 1$, while for $0<\alpha<1$, we have $\lambda(k)\propto \epsilon^{1-2\alpha}$
in the same limit, so the consistency with asymptotic safety requires $\alpha \ge 1/2$.

 We plot the effective gravitational and cosmological constants, $G(\epsilon)$ and $\Lambda(\epsilon)$, as functions of $\epsilon$ for $\alpha=1/2$, $1$ and $2$ in Fig.~\ref{fg:GLambda}.  We can see that $G(\epsilon)$ decreases as $\epsilon$ increases for all 
$\alpha>0$ but faster than $\epsilon^{-1}$ 
only for $\alpha>1$.
Equation~(\ref{eq:effective_bare_stress-energy_tensor})
implies that because of the behavior of $G(\epsilon)$ in the limit of $\epsilon\to \infty$, the contribution of the bare stress-energy tensor to the effective one diverges for $0<\alpha<1$, remain finite for $\alpha=1$ but converges to zero for $\alpha>1$. 
From the asymptotic behavior of $\Lambda(\epsilon)$, we can conclude that only for $\alpha>1$, 
the contribution of the bare matter field is being lost and 
the effective cosmological constant really approaches a positive constant in a high-energy regime.
The above discussion will not depend on the EOS of the fluid. 

\begin{figure}[htbp]
\begin{center}
 \includegraphics[width=0.5\textwidth]{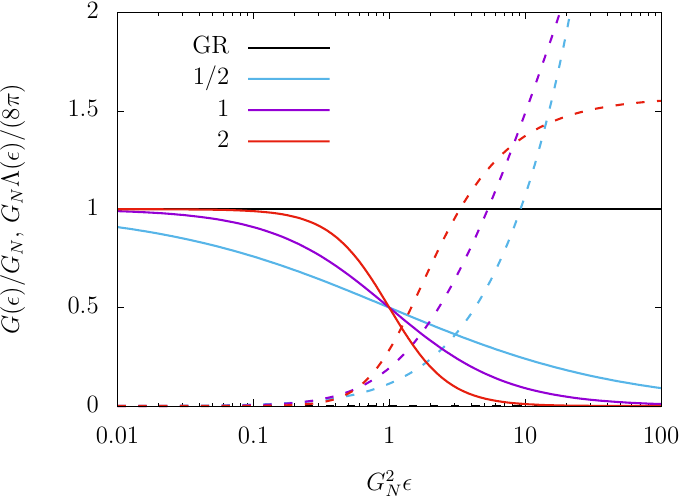}
\end{center}
\caption{The effective gravitational and cosmological constants, $G(\epsilon)$ and $\Lambda(\epsilon)$, are plotted with the solid line and the dashed line, respectively, as functions of $\epsilon$ for $\alpha=1/2$, $1$ and $2$, where $\tilde{\omega}=1$ is fixed. 
Each curve with quantum modification is labeled with the value of $\alpha$. 
\label{fg:GLambda}}
\end{figure}

The behavior of $\epsilon_{\rm eff}$ and $p_{\rm eff}$ as functions of $\epsilon$ is plotted for $\alpha=1/2$, $1$ and $2$ for $w=1/3$ in Fig.~\ref{fg:eeffpeff}.
In this figure, we can see that the asymptotic safety significantly 
slows down the increase in $\epsilon_{\rm eff}$ 
compared to the GR case as $\epsilon$ is increased. We can also see that $\epsilon_{\rm eff}$ still blows up together with 
$\epsilon$ for $\alpha=1/2$ and $\alpha=1$ but approaches a constant value $\epsilon_{{\rm eff}{*}}$
for $\alpha=2$.
In fact, the $\epsilon_{\rm eff}$ and $p_{\rm eff}$ diverge to $+\infty$ and $-\infty$, respectively, as $\epsilon$ increases to $+\infty$ 
for $0 < \alpha \le 1$ but take finite limit values $\epsilon_{{\rm eff}{*}}$ and $p_{{\rm eff}{*}}=-\epsilon_{{\rm eff}{*}}$ for $\alpha > 1$. The divergence of $\epsilon_{\rm eff}$ is resolved and the fluid effectively behaves like a cosmological constant in high-energy limit for $\alpha>1$.

\begin{figure}[htbp]
\begin{center}
\begin{tabular}{cc}
 \includegraphics[width=0.48\textwidth]{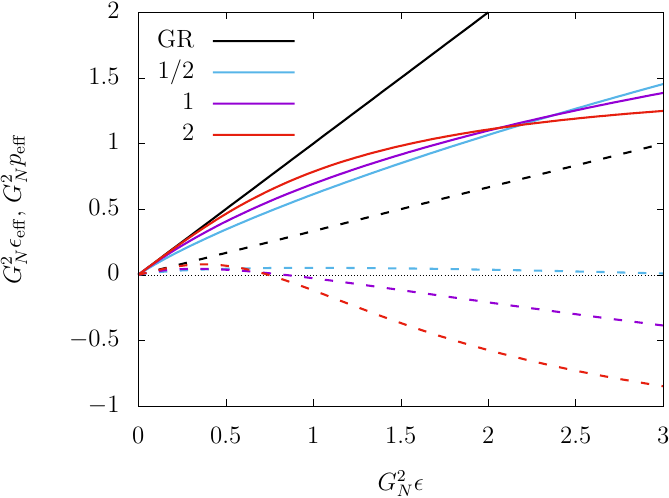} &
  \includegraphics[width=0.48\textwidth]{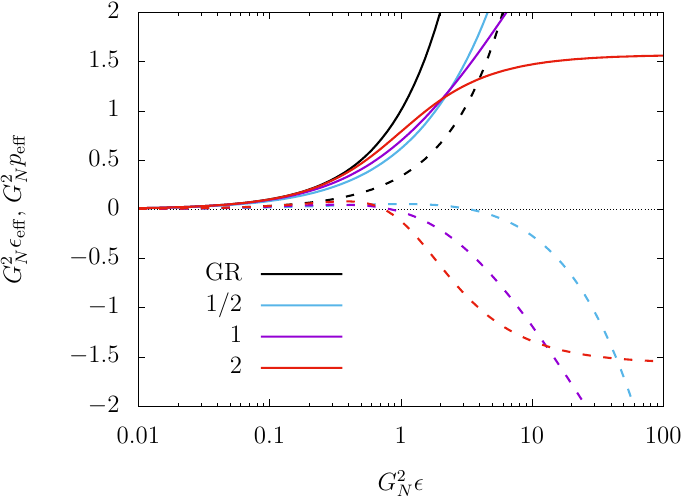} \\
\end{tabular}
\end{center}
\caption{The effective energy density $\epsilon_{\rm eff}$ and $p_{\rm eff}$ are plotted with the solid line and the dashed line, respectively, as functions of $\epsilon$ for $\alpha=1/2$, $1$ and $2$ for the EOS parameter $w=1/3$, where $\tilde{\omega}=1$ is fixed. The left and right panels
show the smaller range of $\epsilon$ in linear scale and the overall range in log scale, respectively. 
Each curve with quantum modification is labeled with the value of $\alpha$. We can see that the increase in $\epsilon_{\rm eff}$ slows down due to the quantum effect compared to the GR case. 
The $\epsilon_{\rm eff} $ and $p_{\rm eff}$ continue increasing to $+\infty$ and decreasing to $-\infty$ 
for $0<\alpha\le 1$ but take finite limit values $\epsilon_{{\rm eff}{*}}$ and $-\epsilon_{{\rm eff}{*}}$ for $\alpha>1$, respectively, as $\epsilon$ increases to $+\infty$.  
\label{fg:eeffpeff}}
\end{figure}

\section{Singularity resolution \label{sec:resolution}}

Here, we will discuss singularity resolution in the FLRW spacetime in the context of gravitational collapse. We should also note that 
the present analysis also immediately applies 
to resolution of big-bang singularities just by taking time reversal.

\subsection{Collapsing FLRW interior}

We assume spherical symmetry for the whole spacetime, homogeneity in the interior and staticity in the exterior. Spherical symmetry and homogeneity directly imply the FLRW spacetime irrespective of gravitational theories.
So, we assume the FLRW metric 
\begin{equation}
 ds^{2} = -d\tau^{2} + a^{2}(\tau) \left( \frac{dr^{2}}{1-Kr^{2}} + r^{2} d\Omega^{2} \right),
\label{eq:FLRW_metric}
\end{equation}
where $d\Omega^{2}$ is the metric on the unit two-sphere and $K$ is constant corresponding to the spatial curvature. The solutions with $K>0$, $K=0$ and $K<0$ are said to be gravitationally bound, marginally bound and unbound, respectively, and all are physically acceptable because $K$ is determined 
just by the initial conditions. 
We further assume the EOS $p = w \epsilon$
for the bare perfect fluid with $w$ being a nonnegative constant. In the current collapse model, the Hamiltonian constraint of the modified Einstein equation reduces to 
\begin{equation}
 \dot{a}^{2} + V(a) + K = 0,
\label{eq:Hubble_eq}
\end{equation}
where $V(a)$ and $\epsilon_{\rm eff}$ are given by 
\begin{eqnarray}
 V(a) = - \frac{a^{2}}{3} 8 \pi G_{N} \epsilon_{\rm eff}(a), \label{eq:V_epsilon_eff} 
 \end{eqnarray}
 where we regard $\epsilon_{\rm eff}$ as a function of $a$.
We should note that $V(a) \le 0$ since $\epsilon_{\rm eff} \ge 0$.

The evolution equations together with the Hamiltonian constraint of the Einstein equation give
\begin{equation}
\frac{\ddot{a}}{a} = - \frac{4 \pi}{3} G_{N}(\epsilon_{\rm eff} + 3 p_{\rm eff}).
\label{eq:addot}
\end{equation}
For the FLRW spacetime, obtaining the solution to the conservation of the number current and substituting it into the bare value $\epsilon = \epsilon(n)$, we find
\begin{eqnarray}
 n = \frac{n_{1}}{a^{3}}, \qquad 
 \epsilon = \frac{\epsilon_{1}}{a^{3(1+w)}},
\label{eq:epsilon_a}
\end{eqnarray}
where $n_{1}$ and $\epsilon_{1}$ are constants of integration.
See Appendix~\ref{sec:matter_conservation} for the conservation law of the bare fluid.

The resulting potentials for $\alpha = 1/2$, $1$ and $2$ for $w = 1/3$ are plotted in Fig.~\ref{fg:potential}. We can see that the modified potential has a negative minimum and approaches $0$ for $\alpha=1$ and $\alpha=2$ or a negative finite value for $\alpha=1/2$ as $a\to 0$, whereas the potential in classical GR diverges to $-\infty$ in the same limit. Then, it follows from Eq.~(\ref{eq:Hubble_eq}) that $a=0$ can be avoided for $K>0$ for $\alpha=1$ and $2$ but is inevitable for $K<0$ even for the modified case, 
whereas it is inevitable irrespective of the value of $K$ in classical GR. The qualitative features do not depend on the EOS as long as $w$ is nonnegative.  
As for the limit of $a\to \infty$, 
because of the behavior of the integral for $0< x \ll 1$, we find 
\begin{eqnarray}
 V(a) \approx -\frac{8 \pi G_{N}}{3} \frac{\epsilon_{1}}{a^{1+3w}}, 
\end{eqnarray}
irrespective of the value of $\alpha$.
\begin{figure}[htbp]
\begin{center}
 \includegraphics[width=0.5\textwidth]{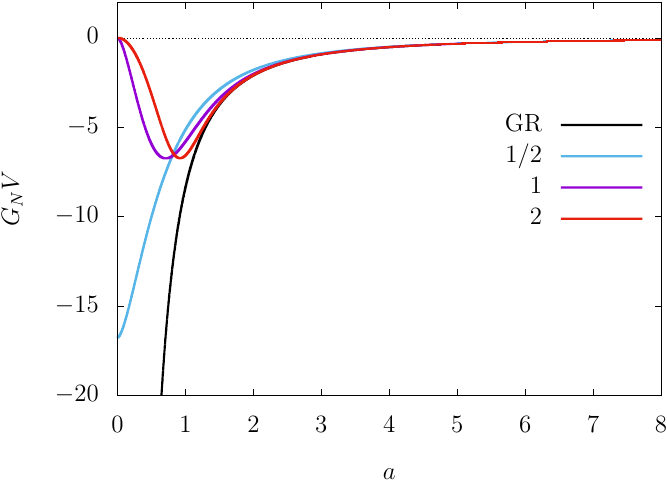}
\end{center}
\caption{The effective potential $V(a)$ for $\alpha=1/2$, $1$ and $2$ for the EOS parameter $w=1/3$, where $\tilde{\omega}=1$ and $G_{N}^{2}\epsilon_{1}=1$ are fixed. 
Each curve with quantum modification is labeled with the value of $\alpha$.
The potential approaches $0$ in the limit $a\to \infty$ for $\alpha>0$, 
while in the limit $a\to 0$ it approaches $0$ for $1$ and $2$ but a finite negative value for $\alpha=1/2$.
We can see that the potential has the only one minimum for $\alpha=1$ and $2$.
\label{fg:potential}}
\end{figure}

The FLRW spacetime has vanishing Weyl curvature and the remaining Riemann tensor is written by the Ricci tensor. All the scalar curvature polynomials are
written solely by $\epsilon_{\rm eff}$ and $p_{\rm eff}$. See, e.g., 
Refs.~\cite{Harada:2018ikn,Harada:2021yul}.
In the FLRW spacetime, Eq.~(\ref{eq:epsilon_a}) follows from the conservation law. 
Therefore, if $a = 0$ is reached in  finite proper time $\tau$, it corresponds to 
curvature singularity for $0<\alpha\le 1$.

\subsection{Dynamics of the scale factor and singularity resolution \label{sec:dynamics}}

Before getting into the analysis of the dynamics of the FLRW solutions and the occurrence of singularities in detail, we present the summary of the results in Table~\ref{table:singularity_resolution_FLRW}. 
As we can see in the table, singularity is  perfectly resolved for $\alpha>1$ irrespective of the spatial curvature $K$, where the complete collapse approaches the de Sitter spacetime.
The case of $\alpha=1$ is special, where singularity
is resolved for $K=0$ by infinite proper time but not resolved for $K<0$. The case of $0<\alpha<1$ 
does not show singularity resolution for $K\le 0$.

\begin{table}
\begin{center}
  \caption{Outcomes of FLRW collapse with $p=w\epsilon$ ($w\ge 0$)}
  \label{table:singularity_resolution_FLRW}
  \centering
  \begin{tabular}{cccccc}
    \hline
    $\alpha $  & $\lim_{\epsilon\to \infty}\Lambda(\epsilon) $ & $K<0$  &  $K=0$ & $K>0$ & Remark  \\
    \hline \hline
    $(1,\infty)$  & $\Lambda_{{*}}$ & de Sitter & de Sitter & ESU or Osc &   \\
    $1$  & $\infty$ & Singularity & $\tau=\infty$ & ESU or Osc & Ref.~\cite{Bonanno:2023rzk}\footnote{The case of $\alpha=1$, $K=0$ and $w=0$ is studied in Ref.~\cite{Bonanno:2023rzk}.} \\
    $(0,1) $ & $\infty$ & Singularity  & Singularity &  
    See Appendix~\ref{sec:alphal1}  & Caveat\footnote{The case of $0<\alpha<1/2$ is inconsistent with asymptotic safety.} \\
    \hline 
    GR & $0$ & Singularity & Singularity & Singularity & E.g. Refs.~\cite{Harada:2018ikn,Harada:2021yul}\\ 
    \hline
  \end{tabular}
  \end{center}
\end{table}

We discuss the behavior of $a(\tau)$ below for the cases $\alpha>1$, $\alpha=1$ and $0<\alpha<1$ separately.

\vspace{0.5cm}
\noindent {1) $\alpha>1$}

In this case, in the limit of $a \to 0$, or $\epsilon = \epsilon_{1}/a^{3(1+w)} \to \infty$, we have
\begin{eqnarray}
 V(a) \approx - \frac{8 \pi G_{N}}{3} a^{2} \tilde{\omega}^{-1/\alpha} G_{N}^{-2} {\mathscr C}_{\alpha} \to 0, 
\end{eqnarray}
Therefore, we deduce that $V(a)$ has a negative minimum 
$V_{\rm min}$ at $a = a_{\rm min}$. This minimum corresponds to the Einstein static universe (ESU) with $p_{\rm eff} = - \epsilon_{\rm eff}/3$ with $K = - V_{\rm min} > 0$ as seen in Eqs.~(\ref{eq:Hubble_eq}) and (\ref{eq:addot})~\cite{Harada:2018ikn, Harada:2021yul}.
We assume that $V(a)$ has the only one minimum.
Since $\epsilon_{\rm eff}$ is a monotonically decreasing function of $a$, $\epsilon_{\rm eff}$ is bounded from above by $\epsilon_{{\rm eff}{*}}$. 
We can also derive 
$p_{\rm eff} \to -\epsilon_{{\rm eff}{*}}$ as $a \to 0$ so that the effective matter field behaves like a cosmological constant. 
Thus, we can conclude that any curvature singularity is completely resolved for any values of $K$ and nonnegative values of $w$. 

The dynamics of the scale factor governed by Eq.~(\ref{eq:Hubble_eq}) is summarized as follows.
For $K>0$, $a$ cannot reach $0$ because $V(a)+K>0$ in the neighborhood of $a=0$. Generically, $a$ oscillates in the interval including $a_{\rm min}$, 
whereas in the exceptional case 
$a$ is constant at $a=a_{\rm min}$ for which we have an ESU. Interestingly, this is stable because a small perturbation gives an oscillating solution around $a=a_{\rm min}$.
For any cases with $K>0$, the effective fluid does not act as a cosmological constant.

For $K=0$, $a$ can reach $0$. We can see that the collapsing solution behaves like
\begin{equation}
 a \approx C \exp\left( -\sqrt{\frac{8\pi G_{N}}{3} \epsilon_{{\rm eff}{*}}} \tau \right)
\label{eq:dS_flat_collapse}
\end{equation}
as $\tau \to \infty$, where $C$ is a constant of integration. So, it takes infinite proper time for $a=0$ to be reached.
In fact, Eq.~(\ref{eq:dS_flat_collapse}) shows a shrinking de Sitter solution in the flat chart, where the timelike geodesic with a constant $r$ finally reaches null infinity in the limit $\tau \to \infty$ as depicted by the orange line in Fig.~\ref{fg:dS_charts}.
This implies the formation of the de Sitter core in the center and this portion is future timelike geodesically complete.

For $K<0$, the potential does not prevent $a$ from reaching $0$ 
in finite proper time. The solution for $a$ is described by 
$a \approx \sqrt{-K}(\tau_{s} - \tau)$
for $a \to 0$, where $\tau = \tau_{s}$ is the time of crossing $a = 0$. 
Although one might regard $\tau = \tau_{s}$ as singularity, it is not. What is really 
happening is the end point of the open chart in the de Sitter spacetime
as depicted by the black filled circle in Fig.~\ref{fg:dS_charts}.
In reality, $\tau=\tau_{s}$ is not singularity but a regular point. 
The spacetime is extended beyond $\tau=\tau_{s}$. 
The interior FLRW region thus {\it vanishes}.
In our case, the exterior and the future of the negative curvature FLRW region is given by a  
regular static metric, which will be discussed later.

\begin{figure}[htbp]
\begin{center}
 \includegraphics[width=0.3\textwidth]{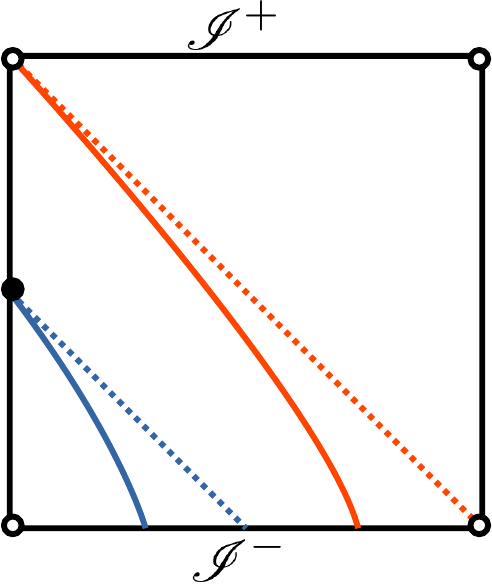}
\end{center}
\caption{Different charts in the de Sitter spacetime. The global chart ($K>0$) covers the whole de Sitter spacetime, which is depicted by the big black square with the coordinate extension through $r=\sqrt{K}\sin\chi$ and $0\le \chi<\pi$. The contracting flat chart ($K=0$) 
covers the left-bottom half below the orange dashed line, while the contracting 
open chart ($K<0$) covers the left-bottom one-eighth below the blue dashed line. 
The world lines with fixed comoving coordinate $r$, which are timelike geodesics, in the flat chart and open chart 
are depicted by the orange solid line and the blue solid line, respectively. 
All world lines with fixed comoving coordinate in the flat chart reach a point depicted by the open circle 
in the future infinity in infinite proper time, 
while those in the open chart terminate at a regular point 
depicted by the filled circle in finite proper time due to coordinate singularity. 
See Refs.~\cite{Sato:1992,Griffiths:2009dfa} for details. \label{fg:dS_charts}}
\end{figure}

\vspace{0.5cm}
\noindent {2) $\alpha=1$}

For this case, we have
\begin{equation}
 V(a) = -\frac{8 \pi G_{N}^{-1}}{3} \tilde{\omega}^{-1} a^{2} \ln\left( 1 + \tilde{\omega} \frac{G_{N}^{2} \epsilon_{1}}{a^{3(1+w)}} \right).
\end{equation}
In this case, $V(a)$ approaches $0$ as $a \to 0$, takes the only one minimum $V = V_{\rm min} < 0$ at $a = a_{\rm min}$ and approaches $0$ as $a \to \infty$. 

The dynamics is summarized as follows.
For $K>0$, $a$ cannot reach $0$ because $V(a) + K > 0$ in the neighborhood of $a = 0$. The solution is either an ESU or an oscillating solution around it.
For $K=0$, it needs infinite proper time for $a=0$ to be reached because the asymptotic solution is given by 
\begin{equation}
 a \approx C \exp\left[ -2 \pi (1+w) \tilde{\omega}^{-1} G_{N}^{-1} \tau^{2} \right],
\end{equation}
as $\tau \to \infty$, where $C$ is a constant of integration. This can be regarded as singularity resolution, which is shown by Bonanno {\it et al.}~\cite{Bonanno:2023rzk}.
For $K<0$, however, $a$ monotonically decreases and reaches $0$ in finite proper time at $\tau = \tau_{s}$ because $V(a) + K \le K < 0$. Since $V(a) \to 0$ as $a \to 0$, $a$ behaves as 
$a \approx \sqrt{-K}(\tau_{s} - \tau)$
for $a \to 0$. In this case, 
$\epsilon_{\rm eff}$ shows divergence,
corresponding to curvature singularity.
Although singularity occurrence is inevitable for unbound collapse, it deserves great attention that curvature strength of the singularity in this case is significantly weak as the scalar curvature polynomials such as ${\cal K}$ only show logarithmic divergence with respect to the affine parameter in an approach of timelike geodesics to the singularity. See Ref.~\cite{Clarke:1994cw} for curvature strength of singularities.

\vspace{0.5cm}
\noindent {3) $0<\alpha<1$}

We delegate the analysis on this case to Appendix~\ref{sec:alphal1} because it needs a special treatment for $K>0$. 
The result is no resolution of singularity for $K\le 0$ as summarized
in Table~\ref{table:singularity_resolution_FLRW}.

\section{Regular black holes, gravastars and uniform cores \label{sec:exterior}}

\subsection{Exterior metric}
Following the prescription in Ref.~\cite{Bonanno:2023rzk}, 
we assume a static exterior region with the metric written in the form
\begin{equation}
 ds^{2} = -f(R) dt^{2} + \frac{1}{f(R)} dR^{2} + R^{2} d\Omega^{2}
\label{eq:f_f-1_metric}
\end{equation}
and require smooth matching by imposing the continuity of both the first and second fundamental forms on the boundary hypersurface 
$r=r_{b}$ between the interior FLRW solution and the exterior. 
If $f(R)>0$ for $0<R<\infty$, there is no horizon, while $f(R) = 0$ at $R = R_{h}$, there is a Killing horizon there.  
As is shown in Appendix~\ref{sec:junction}, it has turned out that the areal radius and the Misner-Sharp mass must be continuous at the matching surface $\Sigma$: $r=r_{b}$ or $R=R_{b}(\tau) = r_{b} a(\tau)$.
Following this procedure, putting 
\begin{equation}
 f(R) = 1 - \frac{2 G_{N} m(R)}{R},
\label{eq:f_M} 
\end{equation}
where $m(R)$ is the Misner-Sharp mass,
the smooth matching suggests 
\begin{equation}
 m(R) = \frac{4 \pi} {3} R^{3} \epsilon_{\rm eff} \left( \frac{R}{r_{b}} \right)
= \frac{4 \pi G_{N}^{-2}}{3} R^{3} \tilde{\omega}^{-1/\alpha} I_{\alpha} \left( \frac{\tilde{\omega}^{1/\alpha} G_{N}^{2} \epsilon_{1} r_{b}^{3(1+w)}}{R^{3(1+w)}} \right),
\label{eq:MR}
\end{equation}
where $\epsilon_{\rm eff} = \epsilon_{\rm eff}(a)$ and the concrete expression for 
$I_{\alpha}(x)$ is given in terms of the Lerch transcendent by Eq.~(\ref{eq:Ialpha_hypergeometric}).~\footnote{It should be noted that the metric functions of regular black holes in some class of quasi-topological theories of gravity are also written in terms of the Lerch transcendent as shown in Ref.~\cite{Frolov:2024hhe}.} Strictly speaking, the junction condition only determines $m(R)$ for the range that the junction surface sweeps, so the above metric should be regarded as analytic continuation if we use it beyond that range.

Irrespective of the value of $\alpha$, we can derive the asymptotic form of $m(R)$ in the limit $R \to \infty$ as
\begin{equation}
 m(R) \approx 4 \pi \frac{\epsilon_{1} r_{b}^{3(1+w)}}{3 R^{3w}}.
\end{equation}
Therefore, if $w \ge 0$, the spacetime is asymptotically flat. 
For $w=0$, the ADM mass $M$ is given by 
$ M = ({4 \pi}/{3}) \epsilon_{1} r_{b}^{3} $, 
while $M = 0$ for $w > 0$. Inspired by the above, we define 
$ M_{w} = ({4 \pi}/{3}) \epsilon_{1} r_{b}^{3(1+w)} $
so that $M_{0}=M$ and 
$ m(R) \approx {M_{w}}/{R^{3w}} $
as $R\to \infty$.
Using this, we can rewrite $m(R)$ as 
\begin{equation}
m(R) 
= \frac{4 \pi G_{N}^{-2}}{3} R^{3} \tilde{\omega}^{-1/\alpha} I_{\alpha} \left( \tilde{\omega}^{1/\alpha} \frac{3 G_{N}^{2}M_{w}}{4 \pi R^{3(1+w)}} \right).
\label{eq:MR_Mw}
\end{equation}
We can show that scalar curvature polynomials are diverging in the limit $R \to 0$ 
for $0 \le \alpha \le 1$, while they are all kept finite for $\alpha>1$. This is
consistent with the analysis of the singularity resolution in the interior region.

We can write the effective gravitational constant $G(\epsilon)$ as a function of $R$ as follows:
\begin{equation}
G = \left. G_{N} \middle/ \left[ 1 + \tilde{\omega} \left( \frac{3 G_{N}^{2} M_{w}}{4 \pi R^{3(1+w)}} \right)^{\alpha} \right] \right. .
\end{equation} 
However, what is more phenomenologically relevant might be the gravitational constant $G_{\rm ex}(R)$ for the exterior defined as 
\begin{equation}
f(R)=1-\frac{2G_{\rm ex}(R)M}{R},
\label{eq:fR}
\end{equation}
so that 
\begin{equation}
G_{\rm ex}(R) = \frac{G_{N} m(R)}{M} = \frac{4 \pi R^{3} \tilde{\omega}^{-1/\alpha}}{3 G_{N} M} I_{\alpha} \left( \tilde{\omega}^{1/\alpha} \frac{3 G_{N}^{2}M}{4 \pi R^{3}} \right)
\end{equation}
follows for $w=0$.
We can recover the classical Schwarzschild metric in the limit $\tilde{\omega} \to 0$ except for $R=0$. 

\subsection{Formation of regular black holes, gravastars, and uniform cores}

For $K>0$, $a$ can remain finite and we are left with a regular homogeneous core that 
is a truncated ESU or an oscillating solution around it. 
We can have the static exterior outside the nonzero finite matching radius $R_{b}=r_{b}a(\tau)$ as discussed above. 
We call these spacetimes uniform cores.
The static uniform core is stable and has no horizon. 

We focus on the most interesting case $\alpha > 1$, where singularity resolution is perfect as we have already shown. For $\alpha > 1$ and $K \le 0$, taking the limit $R \to 0$ in the exterior metric, we can show
\begin{equation}
 m(R) \approx \frac{4 \pi}{3} \epsilon_{{\rm eff}{*}} R^{3},
\end{equation}
where $\epsilon_{{\rm eff}{*}}$ is defined in Eq.~(\ref{eq:eeff0}), and therefore the metric there is given by the de Sitter solution in the static chart such as
\begin{equation}
 ds^{2} \approx -\left( 1 - \frac{8 \pi G_{N}}{3} \epsilon_{{\rm eff}{*}} R^{2} \right) dt^{2}
+ \left( 1 - \frac{8 \pi G_{N}}{3} \epsilon_{{\rm eff}{*}} R^{2} \right)^{-1} dR^{2}
+ R^{2} d\Omega^{2}.
\end{equation}
This implies that the effective cosmological constant near $R=0$ is given by 
$
 \Lambda_{{*}} = 8 \pi G_{N} \epsilon_{{\rm eff}{*}}$,
which is consistent with Eq.~(\ref{eq:Lambda_eff_a>1}). This is nothing but the de Sitter core. The whole metric is given by putting the concrete form of $I_{\alpha}(x)$. The 
$\Lambda_{{*}}$ 
does not depend on the mass of the object.

If the de Sitter core has a horizon, it can describe a regular black hole, while if it does not, it can still describe a compact star with a de Sitter core, or a gravastar. 
As we will see below, the number of positive zeros of $f(R)$ changes as $2$, $1$ and $0$
as $\tilde{\omega}$ is increased from a positive infinitesimal.
If there is only one positive zero, the spacetime describes an extremal black hole and the zero corresponds to an event horizon.
If there are two distinct positive zeros, the spacetime describes a subexremal black hole, where 
the larger and smaller zeros correspond to an outer horizon and an inner horizon, 
respectively, the former of which corresponds to an event horizon.

For simplicity, let us focus on the dust case, where $w=0$.
The radius of antiscreening is found from Eq.~(\ref{eq:MR_Mw}) as
\begin{equation}
 R_{\rm AS} = \left( \tilde{\omega}^{1/\alpha} \frac{3 G_{N}^{2}M}{4\pi} \right)^{1/3} = \left( \frac{6 {\mathscr C}_{\alpha} G_{N} M}{\Lambda_{{*}}} \right)^{1/3}.
\end{equation}
This also gives the radius, where $G$ significantly deviates from $G_{N}$.
The de Sitter core radius $R_{\rm core}$ can be roughly estimated as
$R_{\rm core} \simeq 2 R_{\rm AS}$ and therefore depends on the ADM mass. 
Using this, the condition for the de Sitter core to have a horizon, $R_{\rm core} \lesssim 2 G_{N}M$, reduces to
\begin{equation}
 \tilde{\Omega} := \frac{\tilde{\omega}^{1/\alpha}}{\frac{4\pi}{3} G_{N}M^{2}} \lesssim 1
 \quad {\rm or}\quad  G_{N}^{2}M^{2} \Lambda_{{*}} \gtrsim \frac{3}{4} {\mathscr C}_{\alpha}.
\label{eq:Omega}
\end{equation} 
This can also be written in terms of the critical mass $M_{c}$ as
\begin{equation}
 M \ge M_{c} \simeq \sqrt{\frac{3 \tilde{\omega}^{1/\alpha}}{4 \pi}} m_{P}. 
\end{equation}

Using Eq.~(\ref{eq:Ihalf_I1_I2}), 
the mass function $m(R)$ can be written as
\begin{equation}
 m(R)=\frac{4\pi}{3}R^{3}\tilde{\omega}^{-1/2}G_{N}^{-2}I_{\alpha}\left(\tilde{\omega}^{1/2}\frac{3G_{N}^{2}M}{4\pi R^{3}}\right).
\end{equation}
We can also calculate
\begin{equation}
\frac{G}{G_{N}} = \left[ 1 + \tilde{\omega} \left( \frac{3 G_{N}^{2}M}{4 \pi R^{3}} \right)^{\alpha} \right]^{-1}
= \frac{1}{1 + (R_{\rm AS}/R)^{3\alpha}}
\end{equation} 
and 
\begin{equation}
\frac{G_{\rm ex}}{G_{N}} = \frac{4 \pi}{3 G_{N}^{2} M} R^{3} \tilde{\omega}^{-1/2} 
I_{\alpha}\left( \tilde{\omega}^{1/2} \frac{3 G_{N}^{2}M}{4 \pi R^{3}} \right)
=\left( \frac{R}{R_{\rm AS}} \right)^{3} I_{\alpha}\left(\left( \frac{R}{R_{\rm AS}} \right)^{-3}\right). 
\end{equation}
We plot them for $\alpha=2$ in Fig.~\ref{fg:G_eff}, where we can see that $R_{\rm AS}$ gives the radius of antiscreening of gravity. We can also see that $G$ and $G_{\rm ex}$ are 
different but show very similar behaviors.
\begin{figure}[htbp]
\begin{center}
 \includegraphics[width=0.5\textwidth]{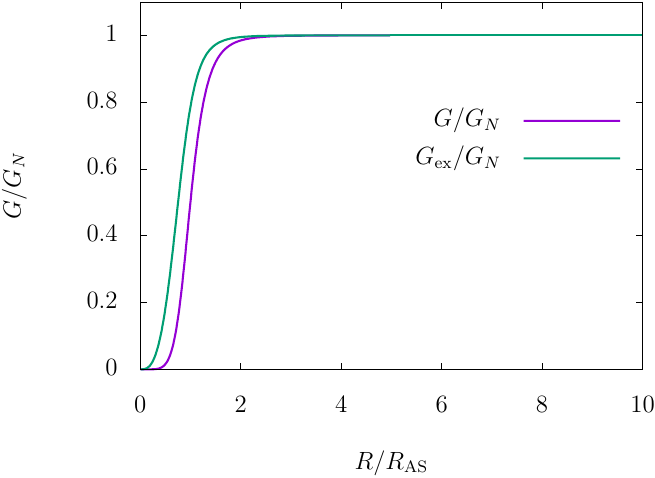}
\end{center}
\caption{The effective gravitational constant $G$ and the gravitational constant $G_{\rm ex}$ for the exterior are plotted for $\alpha=2$ and $w=0$. \label{fg:G_eff}}
\end{figure}

It would be interesting to plot $f(R)$ for different parameter values. 
In this case, we can rewrite $f(R)$ as 
\begin{equation}
 f(R) = 1 - 2 \tilde{\Omega}^{-1} \left( \frac{R}{G_{N}M} \right)^{2} I_{\alpha}\left[ \tilde{\Omega} \left( \frac{R}{G_{N} M} \right)^{-3} \right].
\end{equation}
In Fig.~\ref{fg:fR_alpha}, we plot $f(R)$ for different values of $\tilde{\Omega}$
with the horizontal axis $R$ normalized by half the gravitational radius, $G_{N}M$, for (a) $\alpha=1/2$, (b) $1$, (c) $2$ and (d) $3$.
We should note that only the cases of 
$\alpha>1$, that is, panels (c) and (d) correspond to regular black holes.
We can see that if we fix $M$, the larger the $\tilde{\omega}$ is, the greater the modified effect becomes. 
So, $\tilde{\Omega} \simeq 1$ is the condition that divides a regular black hole and a gravastar.
In fact, we can find the critical values $\tilde{\Omega}_{c} \simeq 2-5$ for $\alpha=1/2, 1, 2$ and $3$. Interestingly, for $\alpha =2$ and $3$, the radius of the event horizon is fairly insensitive to the value of the quantum correction 
as long as it has an event horizon or $\tilde{\Omega}$ is smaller than the critical value $\tilde{\Omega}_{c}$. Using $\tilde{\Omega}_{c}$, we can express the critical mass $M_{c}$ as 
\begin{equation}
M_{c}=\sqrt{\frac{\tilde{\omega}^{1/\alpha}}{\frac{4\pi}{3}G_{N}\tilde{\Omega}_{c}}}=\sqrt{\frac{\tilde{\omega}^{1/\alpha}}{\frac{4\pi}{3}\tilde{\Omega}_{c}}
}m_{P},
\end{equation}
below which the black hole event horizon disappears.
We can also observe that, at least qualitatively, the deviation of the metric functions from the classical Schwarzschild ones in the exterior to the event horizon is larger for the smaller values of $\alpha$.

\begin{figure}[htbp]
\begin{center}
\begin{tabular}{cc}
 \includegraphics[width=0.45\textwidth]{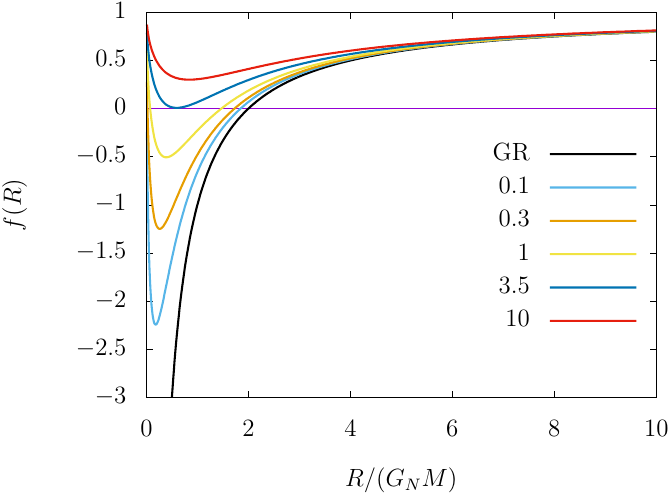} & \includegraphics[width=0.45\textwidth]{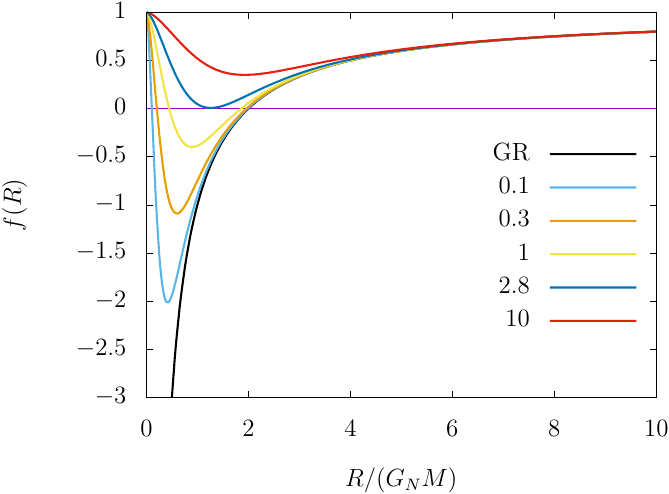} \\
 (a) & (b) \\
 \includegraphics[width=0.45\textwidth]{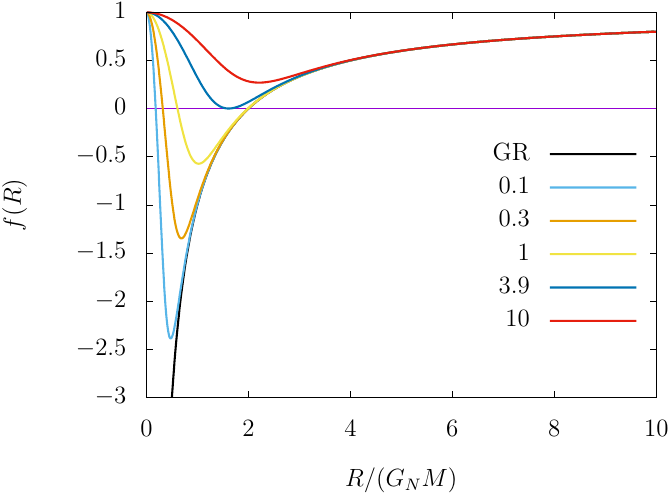} & \includegraphics[width=0.45\textwidth]{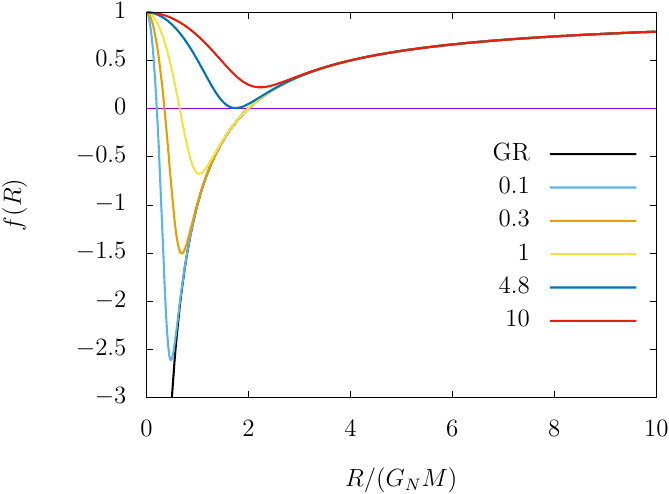} \\
 (c) & (d) 
\end{tabular}
\caption{The function $f(R)$ for different values of 
$\tilde{\Omega}=\tilde{\omega}^{1/\alpha}/[(4\pi/3)G_{N}M^{2}]$ for (a) $\alpha=1/2$, (b) $1$, (c) $2$ and (d) $3$, where $w=0$ is fixed. Each curve is labeled with the value of $\tilde{\Omega}$. We can see that the critical values $\tilde{\Omega}_{c}$ for the disappearance of horizons are approximately given by 
$3.5$. $2.8$, $3.9$ and $4.8$, respectively. Note that only the cases of $\alpha>1$, that is, panels (c) $\alpha=2$ 
and (d) $\alpha=3$ correspond to 
regular black holes.
\label{fg:fR_alpha}}
\end{center}
\end{figure}

\section{Discussion \label{sec:discussion}}

\subsection{Evaporation process of the regular black holes \label{sec:evaporation}}

For the regular black holes discussed above, we can calculate the Hawking temperature $T_{H} = \kappa/(2\pi)$, where 
$
    \kappa = {f'(R_{h+})}/{2}
$
is the surface gravity at the event horizon $R=R_{h+}$. If we fix the theory parameters $\tilde{\omega}$ and $\alpha$ during semiclassical quasistatic evaporation, we expect that the mass of the black hole changes according to  
\begin{equation}
 \frac{dM}{dt} \simeq -4 \pi R_{h+}^{2} \Gamma_{\rm eff} g_{\rm eff} \sigma T_{H}^{4},
\label{eq:mass_loss_rate}
\end{equation}
where $\sigma = \pi^{2}/60$, $g_{\rm eff}$ and $\Gamma_{\rm eff}$ are the Stefan-Boltzmann constant, the effective degrees of freedom and the 
effective gray-body factor, respectively. 

If the mass is large enough so that $M \gg M_{c}$, $T_{H}$ is approximately given by its Schwarzschild value $1/(8\pi M)$ and the evaporation timescale is given by $t_{\rm ev} \simeq G_{N}^{2} M^{3}/(\Gamma_{\rm eff}g_{\rm eff})$. So, the time evolution of $M$ is given by 
\begin{equation}
 M \simeq \left( 1 - \frac{t - t_{i}}{t_{\rm ev}} \right)^{1/3} M_{i},
\end{equation} 
where $M=M_{i}$ is the initial mass at $t=t_{i}$. 

If the mass equals to a critical value $M_{c}$, the horizon becomes degenerate and the Hawking temperature vanishes, where $M_{c}$ is determined by the degeneracy or extremality condition. 
Thus, as the mass decreases to a value very close to the critical one, the 
temperature quickly drops and the evaporation is strongly suppressed. We call this phase late-time evaporation.

To derive the behavior of the 
late-time evaporation, we explicitly write $f$ as a function of both $R$ and $M$, i.e.,  
$f = f(R,M)$. Since $f = f_{,R} = 0$ at $R = R_{c}$ and $M = M_{c}$, we can have the Taylor-series expansion of $f$ at $(R_{c},M_{c})$ as follows:
\begin{equation}
 f(R,M) = f_{,M}(M-M_{c}) + \frac{1}{2} f_{,RR} (R -R_{c})^{2} + f_{,RM}(R - R_{c}) (M - M_{c}) + \frac{1}{2} f_{,MM} (M - M_{c})^{2} + O(3),
\label{eq:Taylor}
\end{equation} 
where the derivatives are evaluated at $(R_{c},M_{c})$ and $O(3)$ denotes cubic and 
higher order terms in powers of $(R-R_{c})$ and $(M-M_{c})$. For $M\simeq M_{c}$ and $M>M_{c}$, we can find the outer and inner horizons $R=R_{h\pm}$ to the lowest order 
as follows:
\begin{equation}
 R_{h\pm}\simeq R_{c}\pm \sqrt{-2f_{,M}(f_{,RR})^{-1}}(M-M_{c})^{1/2}.
\label{eq:near-extremal_horizons}
\end{equation} 
Using Eqs.~(\ref{eq:Taylor}) and~(\ref{eq:near-extremal_horizons}), the surface 
gravity at the outer horizon is calculated to give
\begin{equation}
 \kappa \simeq \sqrt{-\frac{1}{2} f_{,RR} f_{,M}} (M -M_{c})^{1/2}.
\end{equation}
Equation~(\ref{eq:mass_loss_rate}) then can be integrated to give
\begin{equation}
 M - M_{c} \simeq \frac{1}{C (t - t_{t}) + (M_{t} -M_{c})^{-1}}, 
\end{equation}
where 
\begin{equation}
 C = 4 \pi R_{c}^{2} \Gamma_{\rm eff} g_{\rm eff} \sigma 
\frac{(f_{,RR} f_{,M})^{2}}{4 (2 \pi)^{4}} = \frac{1}{480 \pi} \Gamma_{\rm eff} g_{\rm eff} R_{c}^{2}(f_{,RR} f_{,M})^{2}
\end{equation}
and the integration constant is fixed so that 
$M = M_{t}$ at the transition time $t = t_{t}$ to the late-time evaporation. 
The transition mass $M_{t}$ should be taken as
a few times larger than the critical mass $M_{c}$.
 
Thus, after the mass decreases to $M_{t}$ in the early-time evaporation for the timescale 
$t_{\rm ev}$, the black hole undergoes a transition 
to the late-time evaporation, in which the mass asymptotically 
decreases towards $M_{c}$ and the temperature
decreases in proportion to $t^{-1/2}$ for infinitely long time $t\to \infty$.
We can also conclude that neither regular extremal black holes nor gravastars can be obtained as final outcomes of 
Hawking evaporation of subextremal black holes in finite time.
 
It is clear that the above discussion is too simplified 
and in reality the transition should 
occur continuously.
We should also note that 
other possible quantum gravity effects are 
neglected. Further details of the evaporation process will be presented elsewhere~\cite{Harada:2025}.

\subsection{Interpretation of the exterior metric}

We should be cautioned that the exterior metric obtained here is neither a vacuum solution nor a perfect fluid solution in general. In particular, this is not a solution of the effective action
given by Eq.~(\ref{eq:MM_action}). This is an inevitable consequence of the assumption of smooth matching with the static exterior with the metric in the form of 
Eq.~(\ref{eq:f_f-1_metric}), although requiring the continuity of the first and second fundamental forms 
does not give any physically pathological effects about the matching. We can interpret this as that the action~(\ref{eq:MM_action}) that only admits perfect fluid is adequate to describe the homogeneous and isotropic interior but not to describe the exterior region.
From this point of view, we should construct an effective action which can also deal with matter fields that cannot be described by perfect fluid.
Alternatively, we might still seek for a static exterior 
metric containing only perfect fluid governed by the action~(\ref{eq:MM_action}), for which we have to give up the metric form (\ref{eq:f_f-1_metric}).
We can also infer that the assumption of homogeneous interior or smooth matching or static exterior must be broken. 
In fact, even in standard gravitational collapse in classical GR, it is well known that 
smooth matching between the homogeneous interior and the Schwazrschild exterior can be compatible only with a pressure-free fluid, or dust~\cite{Oppenheimer:1939ue}.
Furthermore, if we assume vacuum, it seems that the action~(\ref{eq:MM_action}) will not give any quantum correction.
We leave these problems for future studies.

We should, however, note that 
the result of the current analysis on singularity resolution in the collapse of a homogenous ball 
is very robust, irrespective of the assumption on the exterior region. 
It would also be robust that for $\alpha>1$, we have a regular final outcome, which is 
most plausibly a regular black hole with a de Sitter core, although the detailed
functional form of the exterior metric can be subject to change for different assumptions on the exterior metric and the matching condition.

\section{Concluding remarks \label{sec:conclusion}}

The implications of the results are multifold. First, if singularity resolution is a necessary feature of asymptotically safe gravity, then within the Markov-Mukhanov effective action framework, the effective gravitational constant must behave as in the case of \( \alpha > 1 \). The decrease in the effective gravitational constant for \( 0 < \alpha \leq 1 \) is not fast enough to fully resolve singularities, imposing a strong constraint on singularity resolution phenomenology.  

Second, our study provides a concrete example of a formation scenario for regular black holes and gravastars driven by quantum gravity effects within asymptotically safe gravity. In particular, it is noteworthy that our findings demonstrate the emergence of a de Sitter core as a natural outcome of singularity-free gravitational collapse. In the present setting, for gravitationally unbound collapse or 
with negative spatial curvature, this mechanism is the only means of preventing singularity formation.  

Third, this phenomenon has potential applications to very small black holes that may have formed in the early Universe, known as primordial black holes. For a comprehensive review of primordial black holes and references, see Ref.~\cite{Byrnes:2025tji}. Since primordial black holes can form at extremely high-energy densities and with very small masses, they are more likely to be affected by quantum gravity effects than standard astrophysical black holes. Investigating how the quantum effects studied in this paper influence primordial black hole formation would be of great interest. We speculate that nearly extremal regular black holes in the late stages of Hawking evaporation, with temperatures approaching zero, may play a significant role in dark matter models. In this regards, the possibility of primordial regular black holes and their observational signatures are discussed in Refs.~\cite{Dymnikova:2015yma,Calza:2024fzo,Calza:2024xdh}.

\vspace{1cm}
\noindent
{\bf Note added.} After the preprint version of the present paper was put on the arXiv, another preprint on the same subject appeared on the arXiv~\cite{Hassannejad:2025maw}, which seems to be complementary to the analysis in the present paper.

\acknowledgements

T.H. is grateful to Akihiro Ishibashi,
Takaaki Ishii, Masashi Kimura, Hideki Maeda, Tomoaki Murata, Takuya Takahashi and Norihiro Tanahashi for fruitful discussion.
The authors thank Kazuharu Bamba, Oleg Evnin, Maxim Khlopov, Alex Koek and Fabio Scardigli for their helpful comments.
The work of T.H. was partially supported by JSPS KAKENHI Grants No. JP20H05853 and No. JP24K07027.
The work of C.-M.C. was supported by the National Science and Technology Council of the R.O.C. (Taiwan) under the Grant No. NSTC 113-2112-M-008-027.
The work of R.M. was supported by the National Science and Technology Council of the R.O.C. (Taiwan) under the Grant No. NSTC 113-2811-M-008-046. T.H. expresses gratitude to Department of Physics, National Central University for its hospitality during the initial phase of the current work.

\appendix

\section{(Non-)conservation of the bare matter field \label{sec:matter_conservation}}

\subsection{Nonconservation of the bare matter field}
Let us first review the construction of the action for an isentropic perfect fluid following 
Ref.~\cite{Hawking:1973uf}.
An isentropic perfect fluid is introduced
with a congruence of timelike curves, which is called a fluid flow,
and a number density.
We denote the tangent vector of the fluid flow and the number density by
$W^{\mu}$ and $n$, respectively.
We define a normalized tangent vector along the fluid flow
\begin{equation}
u^{\mu}=\frac{W^{\mu}}{\sqrt{-g_{\alpha\beta}W^{\alpha}W^{\beta}}}
\end{equation}
and require that $j^{\mu}=nu^{\mu}$ satisfies the conservation law
\begin{equation}
\nabla_{\mu}j^{\mu}=0.
\label{eq:number_conservation}
\end{equation}
For the action given by Eq.~(\ref{eq:original_action}),  
we assume that $\epsilon$ is a function of $n$. 
Varying the fluid action with respect to the fluid flow $W^{\mu}$ keeping $j^{\mu}$
conserved, we obtain the Euler equation
\begin{equation}
 (\epsilon+p)u^{\nu}\nabla_{\nu}u^{\mu}=-(g^{\mu\nu}+u^{\mu}u^{\nu})
\nabla_{\nu}p,
\label{eq:Euler_eq}
\end{equation}
where $p$ is given by Eq.~(\ref{eq:1st_law_bare}).
The number conservation (\ref{eq:number_conservation}) can be rewritten 
in terms of $\epsilon$ and $p$ as 
\begin{equation}
u^{\mu}\nabla_{\mu}\epsilon+(\epsilon+p)\nabla_{\mu}u^{\mu}=0,
\label{eq:energy_eq}
\end{equation}
which is interpreted as the energy conservation.
We can derive the stress-energy 
tensor by the variation of the fluid action with respect to the metric $g_{\mu\nu}$ 
keeping $j^{\mu}$ conserved. The resulting form of the stress-energy tensor is given by Eq.~(\ref{eq:bare_matter_stress-energy_tensor}).
Now the conservation law (\ref{eq:bare_matter_conservation}) 
follows from the invariance of the action against the infinitesimal coordinate transformation.  
This is equivalent to (\ref{eq:Euler_eq}) and  
(\ref{eq:energy_eq}).
For a given metric and an EOS, 
we can obtain a solution of these matter equations in terms of $u^{\mu}$ and $\epsilon$.
We can obtain $n$ by directly solving the number conservation (\ref{eq:number_conservation}).

Let us move onto the Markov-Mukhanov action
(\ref{eq:MM_action}), 
where the matter term $16\pi G_{N}\epsilon$ is replaced by $2\chi(\epsilon_{B})\epsilon_{B}$, where the bare quantities are explicitly labeled with the suffix "$B$" in this section. 
We can rewrite Eq.~(\ref{eq:effective_bare_stress-energy_tensor}) as
\begin{eqnarray}
&& 8\pi G_{N}T^{\mu\nu}_{\rm eff}=8\pi G(\epsilon_{B}(n_{\rm eff}))\tilde{T}^{\mu\nu}_{B}-\Lambda(\epsilon_{B}(n_{\rm eff}))g^{\mu\nu}, \\
&& \tilde{T}^{\mu\nu}_{B}:=[\epsilon_{B}(n_{\rm eff})+p_{B}(n_{\rm eff})]u_{\rm eff}^{\mu}u_{\rm eff}^{\nu}+p_{B}(n_{\rm eff})g^{\mu\nu},
\label{eq:tilde_bare_stress-energy_tensor}
\end{eqnarray}
where $n_{\rm eff}$ and $u_{\rm eff}^{\mu}$ are the conserved number density and the unit fluid four-velocity associated with the effective fluid.
Since the effective and bare fluids follow
different EOS's in general as we have seen in Sec.~\ref{sec:MM}, the solutions $u^{\mu}$ of the Euler equation are different from each other in general. 
This is obvious if we are reminded that fluids following different EOS's will have different propagation speeds of sound waves in general. For the solution $u_{\rm eff}^{\mu}$ of the Euler equation for the effective fluid, we can calculate $n_{\rm eff}$ and then $\epsilon_{B}(n_{\rm eff})$ and $p_{B}(n_{\rm eff})$. However, thus constructed 
$\tilde{T}_{B}^{\mu\nu}$ given by Eq.~(\ref{eq:tilde_bare_stress-energy_tensor}) cannot satisfy the conservation law in general. 

On the other hand, if we require the conservation law for the bare fluid
\begin{equation}
T_{B}^{\mu\nu}=[\epsilon_{B}(n_{B})+p_{B}(n_{B})]u_{B}^{\mu}u_{B}^{\nu}+p_{B}(n_{B})g^{\mu\nu},
\label{eq:conserved_bare_stress-energy_tensor}
\end{equation}
we need to realize that the solution $u_{B}^{\mu}$ that satisfies the Euler equation for the bare fluid is different from $u^{\mu}_{\rm eff}$ in general. So, we have to distinguish between $u^{\mu}_{B}$ and $u^{\mu}_{\rm eff}$ and therefore the conserved number densities
$n_{B}$ and $n_{\rm eff}$.
Thus, we can obtain the bare stress-energy tensor (\ref{eq:conserved_bare_stress-energy_tensor}), 
which trivially is conserved by construction.
However, if we are given the modified action (\ref{eq:MM_action}) only, we have no strong reason to define the conserved 
bare stress-energy tensor $T_{B}^{\mu\nu}$
given by Eq.~(\ref{eq:conserved_bare_stress-energy_tensor}).

\subsection{Conservation of the bare matter field in the FLRW spacetime}

Here, we apply the Markov-Mukhanov action to the FLRW spacetime. Since this spacetime has 
high symmetry, we need great care for the application of the discussion in the previous subsection. 
In the FLRW spacetime with the metric (\ref{eq:FLRW_metric}), 
it is clear that
the conservation of the effective stress-energy tensor (\ref{eq:effective_matter_conservation}), or more precisely, the 
Euler equation admits a trivial solution
\begin{equation}
 u_{\rm eff}^{\mu}=\left(\frac{\partial}{\partial \tau}\right)^{\mu}.
 \label{eq:comoving_4velocity}
\end{equation} 
The conservation of $j^{\mu}=n_{\rm eff} u_{\rm eff}^{\mu}$ yields 
\begin{equation}
 n_{\rm eff}\propto \frac{1}{a^{3}}.
 \label{eq:number_density_FLRW}
\end{equation} 
Then, we can calculate $\epsilon_{\rm eff}=\epsilon_{\rm eff}(n_{\rm eff})$ and $p_{\rm eff}=p_{\rm eff}(n_{\rm eff})$ and 
construct $T_{\rm eff}^{\mu\nu}$. We can understand that 
$\nabla_{\mu}T_{\rm eff}^{\mu\nu}=0$ is satisfied from the discussion in the previous subsection.

We should notice, however, that the 
solution~(\ref{eq:comoving_4velocity}) does not depend on the EOS. This means that although the bare fluid follows another EOS than the effective one, both the bare and effective fluids share the same
solution~(\ref{eq:comoving_4velocity})
and (\ref{eq:number_density_FLRW}), i.e.,  
$u^{\mu}_{\rm eff}=u_{B}$ and $n_{\rm eff}=u_{B}$.
In other words, 
using the same $u_{\rm eff}^{\mu}$ and $n_{\rm eff}$, we can find that 
$\tilde{T}^{\mu\nu}_{B}$ defined by Eq.~(\ref{eq:tilde_bare_stress-energy_tensor}) satisfies 
the conservation law
\begin{equation}
\nabla_{\mu}\tilde{T}^{\mu\nu}_{B}=0.
\end{equation}
This is completely due to high symmetry of the FLRW spacetime and therefore accidental 
in this sense. 
This coincidence occurs if the fluid flow $u^{\mu}_{B}$ determined by $\nabla_{\mu}T_{B}^{\mu\nu}=0$ coincides with $u^{\mu}_{\rm eff}$
determined by $\nabla_{\mu}T_{\rm eff}^{\mu\nu}=0$,  
in other words, when $u^{\mu}_{B}=u^{\mu}_{\rm eff}$ holds.

\section{Analysis on the FLRW collapse dynamics for $0<\alpha<1$ \label{sec:alphal1}}

\subsection{$0<\alpha<1$}

In this case, 
the potential is calculated to give
\begin{equation}
 V(a) \approx - \frac{8 \pi G_{N}^{-1}}{3} \frac{1}{1-\alpha} \tilde{\omega}^{-1} (G_{N}^{2} \epsilon_{1})^{1-\alpha} a^{2-3(1-\alpha)(1+w)}
\end{equation}
as $a \to 0$. We put $\beta := 1-(3/2)(1-\alpha)(1+w)$, where $\beta<1$ holds from the assumption. We separately discuss the dynamics for the cases of $\beta>0$, $\beta=0$ and $\beta<0$ below. The conclusion is that we cannot resolve singularity for $K\le 0$ in any case.

For $\beta>0$, $V(a)\to  0$ as $a\to 0$ and there exists a negative minimum 
$V_{\rm min}$ at $a=a_{\rm min}$. Then, 
for $K>0$, $a=0$ cannot be reached. For $K=0$, we have
\begin{equation}
 a \approx \left[ (1-\beta) \sqrt{\frac{8 \pi G_{N}^{-1}}{3} \frac{\tilde{\omega}^{-1}}{1-\alpha}} \right]^{1/(1-\beta)}(\tau_{s}-\tau)^{1/(1-\beta)}
\end{equation}  
as $a \to 0$. Thus, the collapse encounters singularity in finite proper time.
$K<0$ gives singularity in finite proper time at $\tau=\tau_{s}$, where
$
 a \approx \sqrt{-K}(\tau_{s} - \tau)
$
for $a \to  0$.  

For $\beta < 0$, $V(a) \to -\infty$ as $a \to 0$. In this case, singularity occurs in finite proper time for any value of $K$.

For $\beta=0$, $V(a) \to V_{0}<0$ as $a \to 0$, where 
\begin{equation}
 V_{0} = -\frac{8 \pi G_{N}^{-1}}{3} \frac{1}{1-\alpha} \tilde{\omega}^{-1} (G_{N}^{2} \epsilon_{1})^{1-\alpha}.
\end{equation}
For $K + V_{0} > 0$, the collapse will not reach $a = 0$. 
For $K + V_{0} < 0$, singularity occurs if the collapse starts with sufficiently small $a$. 
We will delegate the discussion on an intriguing subclass $\alpha = 1/2$ and $w = 1/3$ to
Appendix~\ref{sec:intriguing_subclass}.

\subsection{Special case: $\alpha=1/2$ and $w=1/3$ \label{sec:intriguing_subclass}}

The case $\alpha=1/2$ and $w=1/3$ seems to be well motivated for a simplistic dimensional argument might imply $\alpha=1/2$ as discussed in Sec.~\ref{sec:Geff} and matter fields with asymptotic freedom should 
behave like radiation at high-energy limit. 
In this case, we obtain 
\begin{eqnarray}
 V(a)&=& -\frac{8\pi G_{N}^{-1}}{3}a^{2}2\tilde{\omega}^{-2}\left[\tilde{\omega}\sqrt{G_{N}^{2}\epsilon}-\ln (1+\tilde{\omega}\sqrt{G_{N}^{2}\epsilon})\right], \\
G_{N}^{2}\epsilon_{\rm eff}&=& 2\tilde{\omega}^{-2}\left[\tilde{\omega}\sqrt{G_{N}^{2}\epsilon}-\ln (1+\tilde{\omega}\sqrt{G_{N}^{2}\epsilon})\right], \\
  \Lambda&=& 8\pi G_{N}^{-1}\tilde{\omega}^{-2}\left\{2\left[\tilde{\omega}\sqrt{G_{N}^{2}\epsilon}-\ln(1+\tilde{\omega}\sqrt{G_{N}^{2}\epsilon})\right]-\tilde{\omega}^{2}\frac{G_{N}^{2}\epsilon}{1+\tilde{\omega}\sqrt{G_{N}^{2}\epsilon}}\right\}.
\end{eqnarray}
The dimensionless cosmological constant is given by 
\begin{equation}
 \lambda(k)=\frac{8\pi \omega}{\tilde{\omega}^{2}}\left\{2\left[1-\frac{\ln (1+\tilde{\omega}\sqrt{G_{N}^{2}\epsilon})}{\tilde{\omega}\sqrt{G_{N}^{2}\epsilon}}\right]-\frac{\tilde{\omega}\sqrt{G_{N}^{2}\epsilon}}{1+\tilde{\omega}\sqrt{G_{N}^{2}\epsilon}}\right\}.
\end{equation}
As $\epsilon$ increases from $0$ to $\infty$, this monotonically increase from $0$ to 
${8\pi \omega}/({\tilde{\omega}^{2}})$.
Thus, $\lambda(k)$ is kept finite in the limit $\epsilon\to \infty$. 

Since $\epsilon=\epsilon_{1}/a^{4}$, we find $\epsilon_{\rm eff}\to \infty$ in the limit of $a\to 0$. We can find that $V(a)$ is a monotonically increasing function with $V(0)=V_{0}$, where
\begin{equation}
 V_{0}=-\frac{8\pi G_{N}^{-1}}{3}2\tilde{\omega}^{-1}\sqrt{G_{N}^{2}\epsilon_{1}}
\end{equation}
and $V(a)\to 0$ as $a\to \infty$.
Therefore, $K> -V_{0}$ is prohibited. 
For $K=-V_{0}$, we cannot set a regular initial condition because only $a=0$ is allowed. 
For $K<-V_{0}$, the collapse encounters singularity in finite proper time. So, there is no singularity resolution.
The collapse encounters singularity in finite proper time for $K<-V_{0}$.
The shape of $V(a)$ for $\alpha=1/2$ and $w=1/3$ is plotted by a cyan curve  in Fig.~\ref{fg:potential}.

\section{Junction conditions \label{sec:junction}}

In this section, we implement the matching between the interior and exterior regions following 
Ref.~\cite{Poisson:2009pwt}.
We use the coordinates $x^{\alpha}$ for the four-dimensional spacetime ${\cal M}$ 
and $y^{a}$ for the three-dimensional timelike or spacelike hypersurface $\Sigma$, 
which is given by $\ell(x^{\alpha})=0$.
We define the unit normal vector $n^{\alpha}$ to $\Sigma$ and the induced metric $h_{ab}$ and the extrinsic curvature $K_{ab}$ on $\Sigma$ as
\begin{eqnarray}
n_{\alpha}= \frac{\epsilon \partial _{\alpha }\ell}{\sqrt{|\partial _{\alpha }\ell \partial^{\alpha}\ell|}};~n^{\alpha}n_{\alpha}=\epsilon=\pm 1,~
 e^{\alpha}_{a}= \frac{\partial x^{\alpha}}{\partial y^{a}}, ~h_{ab}=g_{\alpha\beta}e^{\alpha}_{a}e^{\beta}_{b}, ~ K_{ab}=n_{(\alpha;\beta)}e^{\alpha}_{a}e^{\beta}_{b}. 
\end{eqnarray}
We denote the spacetime regions divided by $\Sigma $ with ${\cal M}_{\pm}$.

We assume that the interior region ${\cal M}_{-}$ is with the FLRW metric given by Eq.~(\ref{eq:FLRW_metric}) and the exterior ${\cal M}_{+}$ is with the general static spherically symmetric metric 
\begin{equation}
 ds^{2}=-e^{\nu(R)}dt^{2}+e^{\lambda(R)}dR^{2}+R^{2}d\Omega^{2}.
\end{equation}
The junction hypersurface $\Sigma $ is written by $r=r_{b}$ in ${\cal M}_{-}$ and $t=T_{b}(\tau)$ and $R=R_{b}(\tau)$ in ${\cal M}_{+}$, where $\tau$ is the proper time. We choose $y^{a}=(\tau,\theta,\phi)$.

In ${\cal M}_{-}$, we have 
\begin{eqnarray}
 n^{-}_{\alpha}&=&\frac{a}{\sqrt{1-Kr_{b}^{2}}}(0,1,0,0),~~e^{-\alpha}_{0}= (1,0,0,0), ~~e^{-\alpha}_{2}= (0,0,1,0), ~~e^{-\alpha}_{3}= (0,0,0,1), \nonumber \\
 h^{-}_{ab}&=& \mbox{diag} (-1,(ar_{b})^{2},(ar_{b})^{2}\sin^{2}\theta),
\end{eqnarray}
and the nonvanishing components of the extrinsic curvature on $\Sigma$ is given by
\begin{eqnarray}
 K^{-}_{00}= 0, ~~K^{-}_{22}= ar_{b}\sqrt{1-Kr_{b}^{2}}, ~~K^{-}_{33}= ar_{b} \sin^{2}\theta  \sqrt{1-Kr_{b}^{2}}.
\end{eqnarray}

In ${\cal M}_{+}$, we have
\begin{eqnarray}
n^{+}_{\alpha}&=& N e^{(\nu+\lambda)/2}(-\dot{R}_{b},\dot{T}_{b},0,0),~~e^{+\alpha}_{0}= (\dot{T}_{b},\dot{R}_{b},0,0), ~~e^{+\alpha}_{2}= (0,0,1,0), ~~
e^{+\alpha}_{3}= (0,0,0,1), \nonumber \\
h^{+}_{ab}&=& \mbox{diag}(-e^{\nu}\dot{T}_{b}^{2}+e^{\lambda}\dot{R}_{b}^{2},R_{b}^{2},R_{b}^{2}\sin^{2}\theta),
\end{eqnarray}
where the dot denotes the ordinary derivative with respect to $\tau$ and 
$N$ is a normalization factor.  

Thus, the continuity of the first fundamental form $[h_{ab}]=0$ requires
\begin{eqnarray}
 && -e^{\nu}\dot{T}_{b}^{2}+e^{\lambda}\dot{R}_{b}^{2}=-1,\label{eq:proper_time}\\
 && R_{b}=ar_{b}.
\label{eq:1st_junction}
\end{eqnarray}
Then, Eq.~(\ref{eq:proper_time}) fixes the normalization constant to $N=1$.

Using the nonvanishing independent components of the Christoffel symbols in ${\cal M}_{+}$, 
\begin{eqnarray}
\Gamma^{0}_{01}&=& \frac{\nu'}{2}, ~~
\Gamma^{1}_{00}= \frac{1}{2}e^{\nu-\lambda}\nu',\quad 
\Gamma^{1}_{11}= \frac{\lambda'}{2}, \quad 
\Gamma^{1}_{22}= -Re^{-\lambda}, \quad 
\Gamma^{1}_{33}=-Re^{-\lambda}\sin^{2}\theta
\nonumber \\
\Gamma^{2}_{12}&=&\frac{1}{R},\quad \Gamma^{2}_{33}=-\sin\theta\cos\theta,\quad
\Gamma^{3}_{13}=\frac{1}{R},\quad \Gamma^{3}_{23}=\cot\theta,
\end{eqnarray}
we obtain the nonvanishing components of the extrinsic curvature 
\begin{eqnarray}
 K^{+}_{00}&=& e^{(\nu+\lambda)/2}(-\ddot{R}_{b}\dot{T}_{b}+\dot{R}_{b}\ddot{T}_{b})
+\frac{1}{2}e^{(\nu-\lambda)/2}[(\nu'-\lambda')e^{\lambda}\dot{R}_{b}^{2}-\nu']\dot{T}_{b}, \nonumber \\
 K^{+}_{22}&=& R_{b} e^{(\nu-\lambda)/2}\dot{T}_{b}, ~~
 K^{+}_{33}= R_{b} e^{(\nu-\lambda)/2}\dot{T}_{b} \sin^{2}\theta.
\end{eqnarray}

Using Eqs.~(\ref{eq:proper_time}) and (\ref{eq:1st_junction}) under the assumption $e^{\nu}=f$ and $e^{\lambda}=1/f$, we can find that
$K_{00}^{+}$ takes a particularly simple form as
\begin{eqnarray}
 K^{+}_{00}= -\frac{2\ddot{R}_{b}+f'}{2f\dot{T}_{b}}, ~~ K^{+}_{22}= Rf\dot{T}_{b}, ~~
 K^{+}_{33}= Rf\dot{T}_{b}\sin^{2}\theta,
\end{eqnarray}
where $\dot{T}_{b}$ is given by 
\begin{equation}
 \dot{T}_{b}=\frac{\sqrt{\dot{R}_{b}^{2}+f}}{f}.
\end{equation}

The continuity of the second fundamental form $[K_{ab}]=0$ together with the above equations requires
\begin{eqnarray}
 2\ddot{R}_{b}+f'&=& 0, \label{eq:2nd_junction_00}\\
 R_{b}\sqrt{\dot{R}_{b}^{2}+f}&=& ar_{b}\sqrt{1-Kr_{b}^{2}}. \label{eq:2nd_junction_22}
\end{eqnarray}
From Eq.~(\ref{eq:2nd_junction_22}) together with Eq.~(\ref{eq:1st_junction}), we find 
\begin{equation}
 \dot{R}_{b}^{2}+f=1-Kr_{b}^{2}.
\label{eq:2nd_junction_22'}
\end{equation}
Differentiating this with respect to $\tau$ gives Eq.~(\ref{eq:2nd_junction_00}). 
From Eq.~(\ref{eq:2nd_junction_22'}) with Eq.~(\ref{eq:f_M}), we can derive
\begin{equation}
 \dot{a}^{2}-\frac{2G_{N}m(ar_{b})}{a r_{b}^{3}}+K=0.
\end{equation}
This is consistent with the Einstein equation in the interior, which is Eq.~(\ref{eq:Hubble_eq}), if and only if 
\begin{equation}
 m(ar_{b})=\frac{4\pi}{3}(ar_{b})^{3}\epsilon_{\rm eff}(a)
~~\mbox{or}~~
 m(R_{b})=\frac{4\pi}{3}R_{b}^{3}\epsilon_{\rm eff}\left(\frac{R_{b}}{r_{b}}\right).
\end{equation}
Thus, the second junction condition determines the function $m(R)$. This is equivalent to the 
continuity of the Misner-Sharp mass at the junction hypersurface $\Sigma$.

\bibliographystyle{apsrev4-1}
\bibliography{ref}

\begin{thebibliography}{45}%
\makeatletter
\providecommand \@ifxundefined [1]{%
 \@ifx{#1\undefined}
}%
\providecommand \@ifnum [1]{%
 \ifnum #1\expandafter \@firstoftwo
 \else \expandafter \@secondoftwo
 \fi
}%
\providecommand \@ifx [1]{%
 \ifx #1\expandafter \@firstoftwo
 \else \expandafter \@secondoftwo
 \fi
}%
\providecommand \natexlab [1]{#1}%
\providecommand \enquote  [1]{``#1''}%
\providecommand \bibnamefont  [1]{#1}%
\providecommand \bibfnamefont [1]{#1}%
\providecommand \citenamefont [1]{#1}%
\providecommand \href@noop [0]{\@secondoftwo}%
\providecommand \href [0]{\begingroup \@sanitize@url \@href}%
\providecommand \@href[1]{\@@startlink{#1}\@@href}%
\providecommand \@@href[1]{\endgroup#1\@@endlink}%
\providecommand \@sanitize@url [0]{\catcode `\\12\catcode `\$12\catcode
  `\&12\catcode `\#12\catcode `\^12\catcode `\_12\catcode `\%12\relax}%
\providecommand \@@startlink[1]{}%
\providecommand \@@endlink[0]{}%
\providecommand \url  [0]{\begingroup\@sanitize@url \@url }%
\providecommand \@url [1]{\endgroup\@href {#1}{\urlprefix }}%
\providecommand \urlprefix  [0]{URL }%
\providecommand \Eprint [0]{\href }%
\providecommand \doibase [0]{http://dx.doi.org/}%
\providecommand \selectlanguage [0]{\@gobble}%
\providecommand \bibinfo  [0]{\@secondoftwo}%
\providecommand \bibfield  [0]{\@secondoftwo}%
\providecommand \translation [1]{[#1]}%
\providecommand \BibitemOpen [0]{}%
\providecommand \bibitemStop [0]{}%
\providecommand \bibitemNoStop [0]{.\EOS\space}%
\providecommand \EOS [0]{\spacefactor3000\relax}%
\providecommand \BibitemShut  [1]{\csname bibitem#1\endcsname}%
\let\auto@bib@innerbib\@empty
\bibitem [{\citenamefont {Reuter}(1998)}]{Reuter:1996cp}%
  \BibitemOpen
  \bibfield  {author} {\bibinfo {author} {\bibfnamefont {M.}~\bibnamefont
  {Reuter}},\ }\href {\doibase 10.1103/PhysRevD.57.971} {\bibfield  {journal}
  {\bibinfo  {journal} {Phys. Rev. D}\ }\textbf {\bibinfo {volume} {57}},\
  \bibinfo {pages} {971} (\bibinfo {year} {1998})},\ \Eprint
  {http://arxiv.org/abs/hep-th/9605030} {arXiv:hep-th/9605030} \BibitemShut
  {NoStop}%
\bibitem [{\citenamefont {Souma}(1999)}]{Souma:1999}%
  \BibitemOpen
  \bibfield  {author} {\bibinfo {author} {\bibfnamefont {W.}~\bibnamefont
  {Souma}},\ }\href {\doibase 10.1143/PTP.102.181} {\bibfield  {journal}
  {\bibinfo  {journal} {Progress of Theoretical Physics}\ }\textbf {\bibinfo
  {volume} {102}},\ \bibinfo {pages} {181} (\bibinfo {year} {1999})},\ \Eprint
  {http://arxiv.org/abs/https://academic.oup.com/ptp/article-pdf/102/1/181/19571317/102-1-181.pdf}
  {https://academic.oup.com/ptp/article-pdf/102/1/181/19571317/102-1-181.pdf}
  \BibitemShut {NoStop}%
\bibitem [{\citenamefont {Percacci}(2017)}]{Percacci:2017book}%
  \BibitemOpen
  \bibfield  {author} {\bibinfo {author} {\bibfnamefont {R.}~\bibnamefont
  {Percacci}},\ }\href {\doibase 10.1142/10369} {\emph {\bibinfo {title} {An
  Introduction to Covariant Quantum Gravity and Asymptotic Safety}}}\ (\bibinfo
   {publisher} {World Scientific},\ \bibinfo {year} {2017})\ \Eprint
  {http://arxiv.org/abs/https://www.worldscientific.com/doi/pdf/10.1142/10369}
  {https://www.worldscientific.com/doi/pdf/10.1142/10369} \BibitemShut
  {NoStop}%
\bibitem [{\citenamefont {Reuter}\ and\ \citenamefont
  {Saueressig}(2019)}]{Reuter:2019byg}%
  \BibitemOpen
  \bibfield  {author} {\bibinfo {author} {\bibfnamefont {M.}~\bibnamefont
  {Reuter}}\ and\ \bibinfo {author} {\bibfnamefont {F.}~\bibnamefont
  {Saueressig}},\ }\href@noop {} {\emph {\bibinfo {title} {{Quantum Gravity and
  the Functional Renormalization Group}: {The Road towards Asymptotic
  Safety}}}}\ (\bibinfo  {publisher} {Cambridge University Press},\ \bibinfo
  {year} {2019})\BibitemShut {NoStop}%
\bibitem [{\citenamefont {Weinberg}(1980)}]{Weinberg:1980gg}%
  \BibitemOpen
  \bibfield  {author} {\bibinfo {author} {\bibfnamefont {S.}~\bibnamefont
  {Weinberg}},\ }\enquote {\bibinfo {title} {{ULTRAVIOLET DIVERGENCES IN
  QUANTUM THEORIES OF GRAVITATION}},}\ in\ \href@noop {} {\emph {\bibinfo
  {booktitle} {{General Relativity}: {An Einstein Centenary Survey}}}}\
  (\bibinfo  {publisher} {{Cambridge University Press}},\ \bibinfo {year}
  {1980})\ pp.\ \bibinfo {pages} {790--831}\BibitemShut {NoStop}%
\bibitem [{\citenamefont {Penrose}(1965)}]{Penrose:1964wq}%
  \BibitemOpen
  \bibfield  {author} {\bibinfo {author} {\bibfnamefont {R.}~\bibnamefont
  {Penrose}},\ }\href {\doibase 10.1103/PhysRevLett.14.57} {\bibfield
  {journal} {\bibinfo  {journal} {Phys. Rev. Lett.}\ }\textbf {\bibinfo
  {volume} {14}},\ \bibinfo {pages} {57} (\bibinfo {year} {1965})}\BibitemShut
  {NoStop}%
\bibitem [{\citenamefont {Hawking}\ and\ \citenamefont
  {Penrose}(1970)}]{Hawking:1970zqf}%
  \BibitemOpen
  \bibfield  {author} {\bibinfo {author} {\bibfnamefont {S.~W.}\ \bibnamefont
  {Hawking}}\ and\ \bibinfo {author} {\bibfnamefont {R.}~\bibnamefont
  {Penrose}},\ }\href {\doibase 10.1098/rspa.1970.0021} {\bibfield  {journal}
  {\bibinfo  {journal} {Proc. Roy. Soc. Lond. A}\ }\textbf {\bibinfo {volume}
  {314}},\ \bibinfo {pages} {529} (\bibinfo {year} {1970})}\BibitemShut
  {NoStop}%
\bibitem [{\citenamefont {Hawking}\ and\ \citenamefont
  {Ellis}(2023)}]{Hawking:1973uf}%
  \BibitemOpen
  \bibfield  {author} {\bibinfo {author} {\bibfnamefont {S.~W.}\ \bibnamefont
  {Hawking}}\ and\ \bibinfo {author} {\bibfnamefont {G.~F.~R.}\ \bibnamefont
  {Ellis}},\ }\href {\doibase 10.1017/9781009253161} {\emph {\bibinfo {title}
  {{The Large Scale Structure of Space-Time}}}},\ Cambridge Monographs on
  Mathematical Physics\ (\bibinfo  {publisher} {Cambridge University Press},\
  \bibinfo {year} {2023})\BibitemShut {NoStop}%
\bibitem [{\citenamefont {Wald}(1984)}]{Wald:1984rg}%
  \BibitemOpen
  \bibfield  {author} {\bibinfo {author} {\bibfnamefont {R.~M.}\ \bibnamefont
  {Wald}},\ }\href {\doibase 10.7208/chicago/9780226870373.001.0001} {\emph
  {\bibinfo {title} {{General Relativity}}}}\ (\bibinfo  {publisher} {Chicago
  Univ. Pr.},\ \bibinfo {address} {Chicago, USA},\ \bibinfo {year}
  {1984})\BibitemShut {NoStop}%
\bibitem [{\citenamefont {Oppenheimer}\ and\ \citenamefont
  {Snyder}(1939)}]{Oppenheimer:1939ue}%
  \BibitemOpen
  \bibfield  {author} {\bibinfo {author} {\bibfnamefont {J.~R.}\ \bibnamefont
  {Oppenheimer}}\ and\ \bibinfo {author} {\bibfnamefont {H.}~\bibnamefont
  {Snyder}},\ }\href {\doibase 10.1103/PhysRev.56.455} {\bibfield  {journal}
  {\bibinfo  {journal} {Phys. Rev.}\ }\textbf {\bibinfo {volume} {56}},\
  \bibinfo {pages} {455} (\bibinfo {year} {1939})}\BibitemShut {NoStop}%
\bibitem [{\citenamefont {Bardeen}(1968)}]{Bardeen:1968}%
  \BibitemOpen
  \bibfield  {author} {\bibinfo {author} {\bibfnamefont {J.}~\bibnamefont
  {Bardeen}},\ }\enquote {\bibinfo {title} {{Nonsingular general relativistic
  gravitational collapse}},}\ in\ \href@noop {} {\emph {\bibinfo {booktitle}
  {Proceedings of the 5th International Conference on Gravitation and the
  Theory of Relativity}}}\ (\bibinfo  {publisher} {Pub. House of Tbilisi
  University},\ \bibinfo {year} {1968})\ p.~\bibinfo {pages} {87}\BibitemShut
  {NoStop}%
\bibitem [{\citenamefont {Hayward}(2006)}]{Hayward:2005gi}%
  \BibitemOpen
  \bibfield  {author} {\bibinfo {author} {\bibfnamefont {S.~A.}\ \bibnamefont
  {Hayward}},\ }\href {\doibase 10.1103/PhysRevLett.96.031103} {\bibfield
  {journal} {\bibinfo  {journal} {Phys. Rev. Lett.}\ }\textbf {\bibinfo
  {volume} {96}},\ \bibinfo {pages} {031103} (\bibinfo {year} {2006})},\
  \Eprint {http://arxiv.org/abs/gr-qc/0506126} {arXiv:gr-qc/0506126}
  \BibitemShut {NoStop}%
\bibitem [{\citenamefont {Maeda}(2022)}]{Maeda:2021jdc}%
  \BibitemOpen
  \bibfield  {author} {\bibinfo {author} {\bibfnamefont {H.}~\bibnamefont
  {Maeda}},\ }\href {\doibase 10.1007/JHEP11(2022)108} {\bibfield  {journal}
  {\bibinfo  {journal} {JHEP}\ }\textbf {\bibinfo {volume} {11}},\ \bibinfo
  {pages} {108} (\bibinfo {year} {2022})},\ \Eprint
  {http://arxiv.org/abs/2107.04791} {arXiv:2107.04791 [gr-qc]} \BibitemShut
  {NoStop}%
\bibitem [{\citenamefont {Bonanno}\ \emph {et~al.}(2017)\citenamefont
  {Bonanno}, \citenamefont {Koch},\ and\ \citenamefont
  {Platania}}]{Bonanno:2016dyv}%
  \BibitemOpen
  \bibfield  {author} {\bibinfo {author} {\bibfnamefont {A.}~\bibnamefont
  {Bonanno}}, \bibinfo {author} {\bibfnamefont {B.}~\bibnamefont {Koch}}, \
  and\ \bibinfo {author} {\bibfnamefont {A.}~\bibnamefont {Platania}},\ }\href
  {\doibase 10.1088/1361-6382/aa6788} {\bibfield  {journal} {\bibinfo
  {journal} {Class. Quant. Grav.}\ }\textbf {\bibinfo {volume} {34}},\ \bibinfo
  {pages} {095012} (\bibinfo {year} {2017})},\ \Eprint
  {http://arxiv.org/abs/1610.05299} {arXiv:1610.05299 [gr-qc]} \BibitemShut
  {NoStop}%
\bibitem [{\citenamefont {Chen}\ \emph {et~al.}(2022)\citenamefont {Chen},
  \citenamefont {Chen}, \citenamefont {Ishibashi}, \citenamefont {Ohta},\ and\
  \citenamefont {Yamaguchi}}]{Chen:2022xjk}%
  \BibitemOpen
  \bibfield  {author} {\bibinfo {author} {\bibfnamefont {C.-M.}\ \bibnamefont
  {Chen}}, \bibinfo {author} {\bibfnamefont {Y.}~\bibnamefont {Chen}}, \bibinfo
  {author} {\bibfnamefont {A.}~\bibnamefont {Ishibashi}}, \bibinfo {author}
  {\bibfnamefont {N.}~\bibnamefont {Ohta}}, \ and\ \bibinfo {author}
  {\bibfnamefont {D.}~\bibnamefont {Yamaguchi}},\ }\href {\doibase
  10.1103/PhysRevD.105.106026} {\bibfield  {journal} {\bibinfo  {journal}
  {Phys. Rev. D}\ }\textbf {\bibinfo {volume} {105}},\ \bibinfo {pages}
  {106026} (\bibinfo {year} {2022})},\ \Eprint
  {http://arxiv.org/abs/2204.09892} {arXiv:2204.09892 [hep-th]} \BibitemShut
  {NoStop}%
\bibitem [{\citenamefont {Chen}\ \emph {et~al.}(2024)\citenamefont {Chen},
  \citenamefont {Chen}, \citenamefont {Ishibashi},\ and\ \citenamefont
  {Ohta}}]{Chen:2023wdg}%
  \BibitemOpen
  \bibfield  {author} {\bibinfo {author} {\bibfnamefont {C.-M.}\ \bibnamefont
  {Chen}}, \bibinfo {author} {\bibfnamefont {Y.}~\bibnamefont {Chen}}, \bibinfo
  {author} {\bibfnamefont {A.}~\bibnamefont {Ishibashi}}, \ and\ \bibinfo
  {author} {\bibfnamefont {N.}~\bibnamefont {Ohta}},\ }\href {\doibase
  10.1016/j.cjph.2024.10.001} {\bibfield  {journal} {\bibinfo  {journal} {Chin.
  J. Phys.}\ }\textbf {\bibinfo {volume} {92}},\ \bibinfo {pages} {766}
  (\bibinfo {year} {2024})},\ \Eprint {http://arxiv.org/abs/2308.16356}
  {arXiv:2308.16356 [hep-th]} \BibitemShut {NoStop}%
\bibitem [{\citenamefont {Chen}\ \emph {et~al.}(2023)\citenamefont {Chen},
  \citenamefont {Chen}, \citenamefont {Ishibashi},\ and\ \citenamefont
  {Ohta}}]{Chen:2023pcv}%
  \BibitemOpen
  \bibfield  {author} {\bibinfo {author} {\bibfnamefont {C.-M.}\ \bibnamefont
  {Chen}}, \bibinfo {author} {\bibfnamefont {Y.}~\bibnamefont {Chen}}, \bibinfo
  {author} {\bibfnamefont {A.}~\bibnamefont {Ishibashi}}, \ and\ \bibinfo
  {author} {\bibfnamefont {N.}~\bibnamefont {Ohta}},\ }\href {\doibase
  10.1088/1361-6382/acfc91} {\bibfield  {journal} {\bibinfo  {journal} {Class.
  Quant. Grav.}\ }\textbf {\bibinfo {volume} {40}},\ \bibinfo {pages} {215007}
  (\bibinfo {year} {2023})},\ \Eprint {http://arxiv.org/abs/2303.04304}
  {arXiv:2303.04304 [hep-th]} \BibitemShut {NoStop}%
\bibitem [{\citenamefont {Hassannejad}\ \emph
  {et~al.}(2025{\natexlab{a}})\citenamefont {Hassannejad}, \citenamefont
  {Lambiase}, \citenamefont {Scardigli},\ and\ \citenamefont
  {Shojai}}]{Hassannejad:2024cbu}%
  \BibitemOpen
  \bibfield  {author} {\bibinfo {author} {\bibfnamefont {R.}~\bibnamefont
  {Hassannejad}}, \bibinfo {author} {\bibfnamefont {G.}~\bibnamefont
  {Lambiase}}, \bibinfo {author} {\bibfnamefont {F.}~\bibnamefont {Scardigli}},
  \ and\ \bibinfo {author} {\bibfnamefont {F.}~\bibnamefont {Shojai}},\ }\href
  {\doibase 10.1103/PhysRevD.111.064069} {\bibfield  {journal} {\bibinfo
  {journal} {Phys. Rev. D}\ }\textbf {\bibinfo {volume} {111}},\ \bibinfo
  {pages} {064069} (\bibinfo {year} {2025}{\natexlab{a}})},\ \Eprint
  {http://arxiv.org/abs/2410.15904} {arXiv:2410.15904 [gr-qc]} \BibitemShut
  {NoStop}%
\bibitem [{\citenamefont {Mazur}\ and\ \citenamefont
  {Mottola}(2004)}]{Mazur:2004fk}%
  \BibitemOpen
  \bibfield  {author} {\bibinfo {author} {\bibfnamefont {P.~O.}\ \bibnamefont
  {Mazur}}\ and\ \bibinfo {author} {\bibfnamefont {E.}~\bibnamefont
  {Mottola}},\ }\href {\doibase 10.1073/pnas.0402717101} {\bibfield  {journal}
  {\bibinfo  {journal} {Proc. Nat. Acad. Sci.}\ }\textbf {\bibinfo {volume}
  {101}},\ \bibinfo {pages} {9545} (\bibinfo {year} {2004})},\ \Eprint
  {http://arxiv.org/abs/gr-qc/0407075} {arXiv:gr-qc/0407075} \BibitemShut
  {NoStop}%
\bibitem [{\citenamefont {Visser}\ and\ \citenamefont
  {Wiltshire}(2004)}]{Visser:2003ge}%
  \BibitemOpen
  \bibfield  {author} {\bibinfo {author} {\bibfnamefont {M.}~\bibnamefont
  {Visser}}\ and\ \bibinfo {author} {\bibfnamefont {D.~L.}\ \bibnamefont
  {Wiltshire}},\ }\href {\doibase 10.1088/0264-9381/21/4/027} {\bibfield
  {journal} {\bibinfo  {journal} {Class. Quant. Grav.}\ }\textbf {\bibinfo
  {volume} {21}},\ \bibinfo {pages} {1135} (\bibinfo {year} {2004})},\ \Eprint
  {http://arxiv.org/abs/gr-qc/0310107} {arXiv:gr-qc/0310107} \BibitemShut
  {NoStop}%
\bibitem [{\citenamefont {Ogawa}\ and\ \citenamefont
  {Ishihara}(2023)}]{Ogawa:2023ive}%
  \BibitemOpen
  \bibfield  {author} {\bibinfo {author} {\bibfnamefont {T.}~\bibnamefont
  {Ogawa}}\ and\ \bibinfo {author} {\bibfnamefont {H.}~\bibnamefont
  {Ishihara}},\ }\href {\doibase 10.1103/PhysRevD.107.L121501} {\bibfield
  {journal} {\bibinfo  {journal} {Phys. Rev. D}\ }\textbf {\bibinfo {volume}
  {107}},\ \bibinfo {pages} {L121501} (\bibinfo {year} {2023})},\ \Eprint
  {http://arxiv.org/abs/2303.07632} {arXiv:2303.07632 [hep-th]} \BibitemShut
  {NoStop}%
\bibitem [{\citenamefont {Ogawa}\ and\ \citenamefont
  {Ishihara}(2024)}]{Ogawa:2024joy}%
  \BibitemOpen
  \bibfield  {author} {\bibinfo {author} {\bibfnamefont {T.}~\bibnamefont
  {Ogawa}}\ and\ \bibinfo {author} {\bibfnamefont {H.}~\bibnamefont
  {Ishihara}},\ }\href {\doibase 10.1103/PhysRevD.110.124003} {\bibfield
  {journal} {\bibinfo  {journal} {Phys. Rev. D}\ }\textbf {\bibinfo {volume}
  {110}},\ \bibinfo {pages} {124003} (\bibinfo {year} {2024})},\ \Eprint
  {http://arxiv.org/abs/2409.07818} {arXiv:2409.07818 [hep-th]} \BibitemShut
  {NoStop}%
\bibitem [{\citenamefont {Reuter}\ and\ \citenamefont
  {Weyer}(2004)}]{Reuter:2004}%
  \BibitemOpen
  \bibfield  {author} {\bibinfo {author} {\bibfnamefont {M.}~\bibnamefont
  {Reuter}}\ and\ \bibinfo {author} {\bibfnamefont {H.}~\bibnamefont {Weyer}},\
  }\href {\doibase 10.1103/PhysRevD.69.104022} {\bibfield  {journal} {\bibinfo
  {journal} {Phys. Rev. D}\ }\textbf {\bibinfo {volume} {69}},\ \bibinfo
  {pages} {104022} (\bibinfo {year} {2004})}\BibitemShut {NoStop}%
\bibitem [{\citenamefont {Markov}\ and\ \citenamefont
  {Mukhanov}(1984)}]{Markov:1984nw}%
  \BibitemOpen
  \bibfield  {author} {\bibinfo {author} {\bibfnamefont {M.~A.}\ \bibnamefont
  {Markov}}\ and\ \bibinfo {author} {\bibfnamefont {V.~F.}\ \bibnamefont
  {Mukhanov}},\ }\href@noop {} {\bibfield  {journal} {\bibinfo  {journal} {JETP
  Lett.}\ }\textbf {\bibinfo {volume} {40}},\ \bibinfo {pages} {1043} (\bibinfo
  {year} {1984})}\BibitemShut {NoStop}%
\bibitem [{\citenamefont {Bonanno}\ \emph {et~al.}(2024)\citenamefont
  {Bonanno}, \citenamefont {Malafarina},\ and\ \citenamefont
  {Panassiti}}]{Bonanno:2023rzk}%
  \BibitemOpen
  \bibfield  {author} {\bibinfo {author} {\bibfnamefont {A.}~\bibnamefont
  {Bonanno}}, \bibinfo {author} {\bibfnamefont {D.}~\bibnamefont {Malafarina}},
  \ and\ \bibinfo {author} {\bibfnamefont {A.}~\bibnamefont {Panassiti}},\
  }\href {\doibase 10.1103/PhysRevLett.132.031401} {\bibfield  {journal}
  {\bibinfo  {journal} {Phys. Rev. Lett.}\ }\textbf {\bibinfo {volume} {132}},\
  \bibinfo {pages} {031401} (\bibinfo {year} {2024})},\ \Eprint
  {http://arxiv.org/abs/2308.10890} {arXiv:2308.10890 [gr-qc]} \BibitemShut
  {NoStop}%
\bibitem [{\citenamefont {Zholdasbek}\ \emph {et~al.}(2024)\citenamefont
  {Zholdasbek}, \citenamefont {Chakrabarty}, \citenamefont {Malafarina},\ and\
  \citenamefont {Bonanno}}]{Zholdasbek:2024pxi}%
  \BibitemOpen
  \bibfield  {author} {\bibinfo {author} {\bibfnamefont {A.}~\bibnamefont
  {Zholdasbek}}, \bibinfo {author} {\bibfnamefont {H.}~\bibnamefont
  {Chakrabarty}}, \bibinfo {author} {\bibfnamefont {D.}~\bibnamefont
  {Malafarina}}, \ and\ \bibinfo {author} {\bibfnamefont {A.}~\bibnamefont
  {Bonanno}},\ }\href@noop {} {\  (\bibinfo {year} {2024})},\ \Eprint
  {http://arxiv.org/abs/2405.02636} {arXiv:2405.02636 [gr-qc]} \BibitemShut
  {NoStop}%
\bibitem [{\citenamefont {Biasi}\ \emph {et~al.}(2022)\citenamefont {Biasi},
  \citenamefont {Evnin},\ and\ \citenamefont {Sypsas}}]{Biasi:2022ktq}%
  \BibitemOpen
  \bibfield  {author} {\bibinfo {author} {\bibfnamefont {A.}~\bibnamefont
  {Biasi}}, \bibinfo {author} {\bibfnamefont {O.}~\bibnamefont {Evnin}}, \ and\
  \bibinfo {author} {\bibfnamefont {S.}~\bibnamefont {Sypsas}},\ }\href
  {\doibase 10.1103/PhysRevLett.129.251104} {\bibfield  {journal} {\bibinfo
  {journal} {Phys. Rev. Lett.}\ }\textbf {\bibinfo {volume} {129}},\ \bibinfo
  {pages} {251104} (\bibinfo {year} {2022})},\ \Eprint
  {http://arxiv.org/abs/2209.06835} {arXiv:2209.06835 [gr-qc]} \BibitemShut
  {NoStop}%
\bibitem [{\citenamefont {Barenboim}\ \emph {et~al.}(2024)\citenamefont
  {Barenboim}, \citenamefont {Frolov},\ and\ \citenamefont
  {Kunstatter}}]{Barenboim:2024dko}%
  \BibitemOpen
  \bibfield  {author} {\bibinfo {author} {\bibfnamefont {J.}~\bibnamefont
  {Barenboim}}, \bibinfo {author} {\bibfnamefont {A.~V.}\ \bibnamefont
  {Frolov}}, \ and\ \bibinfo {author} {\bibfnamefont {G.}~\bibnamefont
  {Kunstatter}},\ }\href {\doibase 10.1103/PhysRevResearch.6.L032055}
  {\bibfield  {journal} {\bibinfo  {journal} {Phys. Rev. Res.}\ }\textbf
  {\bibinfo {volume} {6}},\ \bibinfo {pages} {L032055} (\bibinfo {year}
  {2024})},\ \Eprint {http://arxiv.org/abs/2405.13373} {arXiv:2405.13373
  [gr-qc]} \BibitemShut {NoStop}%
\bibitem [{\citenamefont {Bueno}\ \emph
  {et~al.}(2024{\natexlab{a}})\citenamefont {Bueno}, \citenamefont {Cano},
  \citenamefont {Hennigar},\ and\ \citenamefont {Murcia}}]{Bueno:2024eig}%
  \BibitemOpen
  \bibfield  {author} {\bibinfo {author} {\bibfnamefont {P.}~\bibnamefont
  {Bueno}}, \bibinfo {author} {\bibfnamefont {P.~A.}\ \bibnamefont {Cano}},
  \bibinfo {author} {\bibfnamefont {R.~A.}\ \bibnamefont {Hennigar}}, \ and\
  \bibinfo {author} {\bibfnamefont {A.~J.}\ \bibnamefont {Murcia}},\
  }\href@noop {} {\  (\bibinfo {year} {2024}{\natexlab{a}})},\ \Eprint
  {http://arxiv.org/abs/2412.02742} {arXiv:2412.02742 [gr-qc]} \BibitemShut
  {NoStop}%
\bibitem [{\citenamefont {Bueno}\ \emph
  {et~al.}(2024{\natexlab{b}})\citenamefont {Bueno}, \citenamefont {Cano},
  \citenamefont {Hennigar},\ and\ \citenamefont {Murcia}}]{Bueno:2024zsx}%
  \BibitemOpen
  \bibfield  {author} {\bibinfo {author} {\bibfnamefont {P.}~\bibnamefont
  {Bueno}}, \bibinfo {author} {\bibfnamefont {P.~A.}\ \bibnamefont {Cano}},
  \bibinfo {author} {\bibfnamefont {R.~A.}\ \bibnamefont {Hennigar}}, \ and\
  \bibinfo {author} {\bibfnamefont {A.~J.}\ \bibnamefont {Murcia}},\
  }\href@noop {} {\  (\bibinfo {year} {2024}{\natexlab{b}})},\ \Eprint
  {http://arxiv.org/abs/2412.02740} {arXiv:2412.02740 [gr-qc]} \BibitemShut
  {NoStop}%
\bibitem [{\citenamefont {Bonanno}\ and\ \citenamefont
  {Reuter}(2000)}]{Bonanno:2000ep}%
  \BibitemOpen
  \bibfield  {author} {\bibinfo {author} {\bibfnamefont {A.}~\bibnamefont
  {Bonanno}}\ and\ \bibinfo {author} {\bibfnamefont {M.}~\bibnamefont
  {Reuter}},\ }\href {\doibase 10.1103/PhysRevD.62.043008} {\bibfield
  {journal} {\bibinfo  {journal} {Phys. Rev. D}\ }\textbf {\bibinfo {volume}
  {62}},\ \bibinfo {pages} {043008} (\bibinfo {year} {2000})},\ \Eprint
  {http://arxiv.org/abs/hep-th/0002196} {arXiv:hep-th/0002196} \BibitemShut
  {NoStop}%
\bibitem [{\citenamefont {Hassannejad}\ \emph
  {et~al.}(2025{\natexlab{b}})\citenamefont {Hassannejad}, \citenamefont
  {Shojai},\ and\ \citenamefont {Bamba}}]{Hassannejad:2025maw}%
  \BibitemOpen
  \bibfield  {author} {\bibinfo {author} {\bibfnamefont {R.}~\bibnamefont
  {Hassannejad}}, \bibinfo {author} {\bibfnamefont {F.}~\bibnamefont {Shojai}},
  \ and\ \bibinfo {author} {\bibfnamefont {K.}~\bibnamefont {Bamba}},\
  }\href@noop {} {\  (\bibinfo {year} {2025}{\natexlab{b}})},\ \Eprint
  {http://arxiv.org/abs/2503.23193} {arXiv:2503.23193 [gr-qc]} \BibitemShut
  {NoStop}%
\bibitem [{\citenamefont {Starobinsky}(1982)}]{Starobinsky:1982ee}%
  \BibitemOpen
  \bibfield  {author} {\bibinfo {author} {\bibfnamefont {A.~A.}\ \bibnamefont
  {Starobinsky}},\ }\href {\doibase 10.1016/0370-2693(82)90541-X} {\bibfield
  {journal} {\bibinfo  {journal} {Phys. Lett. B}\ }\textbf {\bibinfo {volume}
  {117}},\ \bibinfo {pages} {175} (\bibinfo {year} {1982})}\BibitemShut
  {NoStop}%
\bibitem [{\citenamefont {Starobinsky}(1986)}]{Starobinsky:1986fx}%
  \BibitemOpen
  \bibfield  {author} {\bibinfo {author} {\bibfnamefont {A.~A.}\ \bibnamefont
  {Starobinsky}},\ }\href {\doibase 10.1007/3-540-16452-9_6} {\bibfield
  {journal} {\bibinfo  {journal} {Lect. Notes Phys.}\ }\textbf {\bibinfo
  {volume} {246}},\ \bibinfo {pages} {107} (\bibinfo {year}
  {1986})}\BibitemShut {NoStop}%
\bibitem [{\citenamefont {Harada}\ \emph {et~al.}(2018)\citenamefont {Harada},
  \citenamefont {Carr},\ and\ \citenamefont {Igata}}]{Harada:2018ikn}%
  \BibitemOpen
  \bibfield  {author} {\bibinfo {author} {\bibfnamefont {T.}~\bibnamefont
  {Harada}}, \bibinfo {author} {\bibfnamefont {B.~J.}\ \bibnamefont {Carr}}, \
  and\ \bibinfo {author} {\bibfnamefont {T.}~\bibnamefont {Igata}},\ }\href
  {\doibase 10.1088/1361-6382/aab99f} {\bibfield  {journal} {\bibinfo
  {journal} {Class. Quant. Grav.}\ }\textbf {\bibinfo {volume} {35}},\ \bibinfo
  {pages} {105011} (\bibinfo {year} {2018})},\ \Eprint
  {http://arxiv.org/abs/1801.01966} {arXiv:1801.01966 [gr-qc]} \BibitemShut
  {NoStop}%
\bibitem [{\citenamefont {Harada}\ \emph {et~al.}(2022)\citenamefont {Harada},
  \citenamefont {Igata}, \citenamefont {Sato},\ and\ \citenamefont
  {Carr}}]{Harada:2021yul}%
  \BibitemOpen
  \bibfield  {author} {\bibinfo {author} {\bibfnamefont {T.}~\bibnamefont
  {Harada}}, \bibinfo {author} {\bibfnamefont {T.}~\bibnamefont {Igata}},
  \bibinfo {author} {\bibfnamefont {T.}~\bibnamefont {Sato}}, \ and\ \bibinfo
  {author} {\bibfnamefont {B.}~\bibnamefont {Carr}},\ }\href {\doibase
  10.1088/1361-6382/ac776e} {\bibfield  {journal} {\bibinfo  {journal} {Class.
  Quant. Grav.}\ }\textbf {\bibinfo {volume} {39}},\ \bibinfo {pages} {145008}
  (\bibinfo {year} {2022})},\ \Eprint {http://arxiv.org/abs/2110.13421}
  {arXiv:2110.13421 [gr-qc]} \BibitemShut {NoStop}%
\bibitem [{\citenamefont {Sato}\ and\ \citenamefont
  {Kodama}(1992)}]{Sato:1992}%
  \BibitemOpen
  \bibfield  {author} {\bibinfo {author} {\bibfnamefont {H.}~\bibnamefont
  {Sato}}\ and\ \bibinfo {author} {\bibfnamefont {H.}~\bibnamefont {Kodama}},\
  }\href@noop {} {\emph {\bibinfo {title} {{Ippansoutaiseiriron (In
  Japanese)}}}}\ (\bibinfo  {publisher} {Iwanami Shoten, Publishers},\ \bibinfo
  {address} {Tokyo, Japan},\ \bibinfo {year} {1992})\BibitemShut {NoStop}%
\bibitem [{\citenamefont {Griffiths}\ and\ \citenamefont
  {Podolsky}(2009)}]{Griffiths:2009dfa}%
  \BibitemOpen
  \bibfield  {author} {\bibinfo {author} {\bibfnamefont {J.~B.}\ \bibnamefont
  {Griffiths}}\ and\ \bibinfo {author} {\bibfnamefont {J.}~\bibnamefont
  {Podolsky}},\ }\href {\doibase 10.1017/CBO9780511635397} {\emph {\bibinfo
  {title} {{Exact Space-Times in Einstein's General Relativity}}}},\ Cambridge
  Monographs on Mathematical Physics\ (\bibinfo  {publisher} {Cambridge
  University Press},\ \bibinfo {address} {Cambridge},\ \bibinfo {year}
  {2009})\BibitemShut {NoStop}%
\bibitem [{\citenamefont {Clarke}(1994)}]{Clarke:1994cw}%
  \BibitemOpen
  \bibfield  {author} {\bibinfo {author} {\bibfnamefont {C.~J.~S.}\
  \bibnamefont {Clarke}},\ }\href@noop {} {\emph {\bibinfo {title} {{The
  Analysis of space-time singularities}}}},\ Cambridge Lecture Notes in
  Physics\ (\bibinfo  {publisher} {Cambridge Univ. Press},\ \bibinfo {address}
  {Cambridge, UK},\ \bibinfo {year} {1994})\BibitemShut {NoStop}%
\bibitem [{\citenamefont {Frolov}\ \emph {et~al.}(2025)\citenamefont {Frolov},
  \citenamefont {Koek}, \citenamefont {Soto},\ and\ \citenamefont
  {Zelnikov}}]{Frolov:2024hhe}%
  \BibitemOpen
  \bibfield  {author} {\bibinfo {author} {\bibfnamefont {V.~P.}\ \bibnamefont
  {Frolov}}, \bibinfo {author} {\bibfnamefont {A.}~\bibnamefont {Koek}},
  \bibinfo {author} {\bibfnamefont {J.~P.}\ \bibnamefont {Soto}}, \ and\
  \bibinfo {author} {\bibfnamefont {A.}~\bibnamefont {Zelnikov}},\ }\href
  {\doibase 10.1103/PhysRevD.111.044034} {\bibfield  {journal} {\bibinfo
  {journal} {Phys. Rev. D}\ }\textbf {\bibinfo {volume} {111}},\ \bibinfo
  {pages} {044034} (\bibinfo {year} {2025})},\ \Eprint
  {http://arxiv.org/abs/2411.16050} {arXiv:2411.16050 [gr-qc]} \BibitemShut
  {NoStop}%
\bibitem [{\citenamefont {Byrnes}\ \emph {et~al.}(2025)\citenamefont {Byrnes},
  \citenamefont {Franciolini}, \citenamefont {Harada}, \citenamefont {Pani},\
  and\ \citenamefont {Sasaki}}]{Byrnes:2025tji}%
  \BibitemOpen
  \bibinfo {editor} {\bibfnamefont {C.}~\bibnamefont {Byrnes}}, \bibinfo
  {editor} {\bibfnamefont {G.}~\bibnamefont {Franciolini}}, \bibinfo {editor}
  {\bibfnamefont {T.}~\bibnamefont {Harada}}, \bibinfo {editor} {\bibfnamefont
  {P.}~\bibnamefont {Pani}}, \ and\ \bibinfo {editor} {\bibfnamefont
  {M.}~\bibnamefont {Sasaki}},\ eds.,\ \href@noop {} {\emph {\bibinfo {title}
  {{Primordial Black Holes}}}},\ Springer Series in Astrophysics and Cosmology\
  (\bibinfo  {publisher} {Springer},\ \bibinfo {year} {2025})\BibitemShut
  {NoStop}%
\bibitem [{\citenamefont {Dymnikova}\ and\ \citenamefont
  {Khlopov}(2015)}]{Dymnikova:2015yma}%
  \BibitemOpen
  \bibfield  {author} {\bibinfo {author} {\bibfnamefont {I.}~\bibnamefont
  {Dymnikova}}\ and\ \bibinfo {author} {\bibfnamefont {M.}~\bibnamefont
  {Khlopov}},\ }\href {\doibase 10.1142/S0218271815450029} {\bibfield
  {journal} {\bibinfo  {journal} {Int. J. Mod. Phys. D}\ }\textbf {\bibinfo
  {volume} {24}},\ \bibinfo {pages} {1545002} (\bibinfo {year} {2015})},\
  \Eprint {http://arxiv.org/abs/1510.01351} {arXiv:1510.01351 [gr-qc]}
  \BibitemShut {NoStop}%
\bibitem [{\citenamefont {Calz\`a}\ \emph
  {et~al.}(2025{\natexlab{a}})\citenamefont {Calz\`a}, \citenamefont
  {Pedrotti},\ and\ \citenamefont {Vagnozzi}}]{Calza:2024fzo}%
  \BibitemOpen
  \bibfield  {author} {\bibinfo {author} {\bibfnamefont {M.}~\bibnamefont
  {Calz\`a}}, \bibinfo {author} {\bibfnamefont {D.}~\bibnamefont {Pedrotti}}, \
  and\ \bibinfo {author} {\bibfnamefont {S.}~\bibnamefont {Vagnozzi}},\ }\href
  {\doibase 10.1103/PhysRevD.111.024009} {\bibfield  {journal} {\bibinfo
  {journal} {Phys. Rev. D}\ }\textbf {\bibinfo {volume} {111}},\ \bibinfo
  {pages} {024009} (\bibinfo {year} {2025}{\natexlab{a}})},\ \Eprint
  {http://arxiv.org/abs/2409.02804} {arXiv:2409.02804 [gr-qc]} \BibitemShut
  {NoStop}%
\bibitem [{\citenamefont {Calz\`a}\ \emph
  {et~al.}(2025{\natexlab{b}})\citenamefont {Calz\`a}, \citenamefont
  {Pedrotti},\ and\ \citenamefont {Vagnozzi}}]{Calza:2024xdh}%
  \BibitemOpen
  \bibfield  {author} {\bibinfo {author} {\bibfnamefont {M.}~\bibnamefont
  {Calz\`a}}, \bibinfo {author} {\bibfnamefont {D.}~\bibnamefont {Pedrotti}}, \
  and\ \bibinfo {author} {\bibfnamefont {S.}~\bibnamefont {Vagnozzi}},\ }\href
  {\doibase 10.1103/PhysRevD.111.024010} {\bibfield  {journal} {\bibinfo
  {journal} {Phys. Rev. D}\ }\textbf {\bibinfo {volume} {111}},\ \bibinfo
  {pages} {024010} (\bibinfo {year} {2025}{\natexlab{b}})},\ \Eprint
  {http://arxiv.org/abs/2409.02807} {arXiv:2409.02807 [gr-qc]} \BibitemShut
  {NoStop}%
\bibitem [{\citenamefont {Poisson}(2009)}]{Poisson:2009pwt}%
  \BibitemOpen
  \bibfield  {author} {\bibinfo {author} {\bibfnamefont {E.}~\bibnamefont
  {Poisson}},\ }\href {\doibase 10.1017/CBO9780511606601} {\emph {\bibinfo
  {title} {{A Relativist's Toolkit: The Mathematics of Black-Hole
  Mechanics}}}}\ (\bibinfo  {publisher} {Cambridge University Press},\ \bibinfo
  {year} {2009})\BibitemShut {NoStop}%
\end{thebibliography}%

\end{document}